\def\equationautorefname~#1\null{Equation (#1)\null}
\def\sectionautorefname~#1\null{Section #1\null}
\def\subsectionautorefname~#1\null{Section #1\null}
\def\subsubsectionautorefname~#1\null{Section #1\null}
\def\figureautorefname~#1\null{Figure #1\null}

\documentclass[twocolumn]{aastex63}
\usepackage{lineno}
\usepackage{amsmath}
\usepackage{xparse}
\usepackage{hyperref}
\usepackage{longtable}
\usepackage[flushleft]{threeparttable}
\usepackage[titletoc,title]{appendix}
 \usepackage{multirow}
\usepackage{dcolumn}
\usepackage{bm}
\usepackage{tablefootnote}

\newcommand*{\ar}{\autoref}
\newcommand*{\ct}{\citet}
\newcommand*{\ctp}{\citep}
\newcommand*{\cta}{\citealp}

\received{}
\revised{}
\accepted{\today}
\graphicspath{{./}{figures/}}

\begin{document}

\title{Diversity of early kilonova with the realistic opacities of highly ionized heavy elements}
\correspondingauthor{Smaranika Banerjee}
\email{smaranika.banerjee@astro.su.se}

\author[0000-0002-0786-7307]{Smaranika Banerjee}
\affiliation{Astronomical Institute, Tohoku University, Aoba, Sendai 980-8578, Japan}
\affiliation{Stockholm University, Roslagstullsbacken 21, 114 21 Stockholm, Sweden}

\author[0000-0001-8253-6850]{Masaomi Tanaka}
\affiliation{Astronomical Institute, Tohoku University, Aoba, Sendai 980-8578, Japan}
\affiliation{Division for the Establishment of Frontier Sciences, Organization for Advanced Studies,
Tohoku University, Sendai 980-8577, Japan}

\author[0000-0002-5302-073X]{Daiji Kato}
\affiliation{National Institute for Fusion Science, National Institutes of Natural Sciences, Oroshi-cho, Toki, Gifu 509-5292, Japan}
\affiliation{Department of Advanced Energy Engineering Science, Kyushu University, Kasuga, Fukuoka 816-8580, Japan}

\author[0000-0003-0039-1163]{Gediminas Gaigalas}
\affiliation{Institute of Theoretical Physics and Astronomy, Vilnius University, Saul\.etekio av. 3, LT-10257 Vilnius, Lithuania}

\begin{abstract}

We investigate the early ($t < 1$ day) kilonova from the neutron star merger 
by deriving atomic opacities for all the elements from La to Ra ($Z = 57 - 88$) ionized to the states V - XI.
The opacities at high temperatures for the elements with open $f$-shells (e.g., lanthanides) are exceptionally high,
reaching $\kappa_{\rm exp}\,\sim 10^4 \,\rm{cm^{2}\,g^{-1}}$ at $\lambda \le 1000\, \rm \AA$ at $T\sim 70,000$ K, 
whereas, the opacities at the same temperature and wavelengths for the elements with the open $d$-, $p$-, and $s$-shells reach
$\kappa_{\rm exp}\,\sim 1\,\rm{cm^{2}\,g^{-1}}$, $0.1\,\rm{cm^{2}\,g^{-1}}$, and $0.01 \,\rm{cm^{2}\,g^{-1}}$, respectively.
Using the new opacity dataset, we derive the early kilonovae for various compositions and density structures expected for neutron star merger ejecta.
The bolometric luminosity for the lanthanide-rich ejecta shows distinct signatures and is fainter than that for the lanthanide-free ejecta.
The early luminosity is suppressed by the presence of a thin outer layer, agreeing with the results of \ct{kasen17} and \ct{Smaranikab20}.
The early brightness in \textit{Swift} UVOT filters and in the optical $g$-, $r$-, $i$-, $z$-filters for a source at 100 Mpc are
$\sim 22 - 20$ mag and $\sim 21 - 19$ mag, respectively, at $t\,\sim 0.1$ days. 
Such kilonovae are ideal targets for the upcoming UV satellites, such as ULTRASAT, UVEX, and DORADO, and the upcoming surveys, e.g., Vera Rubin Observatory.
We suggest the gray opacities to reproduce the bolometric light curves with and without lanthanides
are $\sim 1 - 20\,\rm{cm^{2}\,g^{-1}}$ and $\sim 0.8 - 1\,\rm{cm^{2}\,g^{-1}}$.

\end{abstract}


\section{Introduction} \label{sec:intr}

Binary neutron star mergers have long been hypothesized to be one of the most plausible sites for heavy element synthesis.
In the neutron-rich material ejected after the neutron star merger, heavy ($Z > 26$) elements are synthesized via rapid neutron capture
($r$-process, e.g., \cta{Lattimer74, Eichler89, Freiburghaus99, Korobkin12, Wanajo14}).
Radioactive decay of freshly synthesized heavy elements produces a transient in the Ultraviolet-Optical-Infrared (UVOIR) range,
called a kilonova (e.g., \citealt{Li98, Kulkarni05, Metzger10}).
The recent detection of the kilonova AT2017gfo from the neutron star merger (e.g., \cta{Coulter17, Yang17, Valenti17})
by the follow-up observation of the gravitational wave signal GW170817 \citep{Abbott17a}
has considerably progressed our understanding on the origin of heavy elements.

Several efforts have been made to model the kilonova AT2017gfo,
which evolved from UV and optical to near-infrared (NIR) in the timescale of about $\sim$ a week 
(e.g., \cta{Coulter17, Yang17,Valenti17, Cowperthwaite17, Smartt17, Drout17, Utsumi17}).
The light curves at late times ($t > 1$ day, hereafter $t$ denotes time after the merger)
are well explained by the radioactive decay of the heavy elements or kilonova
(e.g., \cta{kasen17, Tanaka17, Shibata17, Perego17, Rosswog18, Kawaguchi18}).
Nonetheless, the source of the emission at early time ($t<1$ day) had been a matter of debate (see \cta{arcavi18}),
since it was not clear whether the radioactive heating is enough to explain the early kilonova (e.g., \cta{Villar17, Waxman18})
or other heating source such as the cooling from the shocked ejecta formed by the jet-ejecta interaction (e.g., \cta{Kasliwal17, Piro18})
or $\beta$ decays of free neutrons \ctp{Metzger15, Gottlieb20} or central engine activity (e.g., \cta{Metzger08, Yu13, Metzger14, Metzger18}) is necessary.
However, now the emission mechanism of kilonova AT2017gfo at $t< 1$ day has mostly reached the consensus to have the same source as the late time,
i.e., kilonova powered by radioactive heating can explain the early light curve (e.g., \cta{Smaranikab20, Klion21}).

Kilonova light curves strongly depend on the bound-bound opacity of the $r$-process elements, which requires the atomic data
(e.g., \cta{kasen13, Tanaka13, Fontes15, Fontes20, Wollaeger17, Tanaka18, Tanaka20a}).
However, such atomic data were largely unavailable (see \cta{NIST20}).
In the past decade, calculations of the atomic opacity have been the intense focus of research
(e.g., \cta{kasen13, Tanaka13, Fontes15, Wollaeger17, Tanaka18, Gaigalas19, Tanaka20a, Fontes20, Rynkun21, Fontes23, Flors23}),
although all of these studies focus on the relatively lower ionized elements, maximum up to third
(or IV in spectroscopic notation; hereafter used to describe the ionization), which are important for the kilonova at late times ($t > 1$ day).

The atomic opacity at an early time ($t < 1$ day) is determined by the highly ionized heavy elements since the neutron star merger ejecta are highly ionized at an early time.
For the AT2017gfo-like binary neutron star merger with an ejecta mass of $M_{\rm ej}\sim 0.05M_{\odot}$ \ctp{Kasliwal17, kasen17, Waxman18, Smaranikab20},
moving with an average velocity of $v_{\rm avg} \sim 0.1c$ (e.g., \cta{Tanaka17}), the maximum ionization at around $t\sim 0.1$ days is $\sim$ XI,
owing to the high temperature of ejecta $T\,\sim\,10^5$ K \ctp{Smaranikab20, Klion21}.
The atomic opacity calculations for the highly ionized (V - XI) elements has been started relatively recently (e.g., \cta{Smaranikab20, Smaranikab22}).
However, the calculations are limited to the lighter $r$-process elements (Ca - Ba, $Z = 20 - 56$, \cta{Smaranikab20})
and selected rare earth elements ($Z = 57 - 71$), lanthanides (see \cta{Smaranikab22} for atomic opacities of Nd, $Z = 60$; Sm, $Z = 62$; and Eu, $Z = 63$;
also see \cta{Carvajal_Gallego22a, Carvajal_Gallego22b, Carvajal_Gallego23a, Carvajal_Gallego23b} for the atomic opacities for La - Sm, $Z = 57 - 61$,
ionized to states V - X; \cta{Maison22} for Lu ionized to V).
The atomic opacity for many of the lanthanides, such as, the lanthanides with $Z = 64 - 71$, and the post-lanthanide $r$-process elements
($Z >71$, e.g., \cta{Wanajo14}) are yet to be included.

Calculation of the atomic opacity for all the elements are required to understand kilonova from different viewing angles.
This is because in neutron star mergers, different elements are inhomogeneously distributed (see \cta{Shibata19}),
which introduces angle dependence in light curves due to the different opacity for different elements.
For example, if the lanthanides are present in the ejecta, then it causes the opacity to increase dramatically
(e.g., \cta{kasen13, Tanaka18, Fontes20, Tanaka20a}), affecting the light curves.
Such lanthanides (also the heavy $r$-process elements $Z >71$) are expected to be mainly distributed in the equatorial region by the dynamical mass ejection.
On the other hand, the lighter $r$-process elements ($Z\le$ 56) are expected to be distributed in the shocked dynamical ejecta in the polar direction 
or in the relatively isotropic outflow from the accretion disk
(e.g., \cta{Bauswein13, Just15, Sekiguchi15, Just22, Metzger14, Miller19, Fernandez14, Perego14, Lippuner17, Fujibayashi18, Fujibayashi20a, Fujibayashi20b}).

Understanding the kilonova from different directions starting from an early time is important to extract the abundances from observations.
The fourth GW observing Run (O4), which starts in the later half of 2023,
is expected to make several joint detections of GW and kilonova per calender year (e.g., \cta{Colombo22}).
The detected events might have viewing angles different from each others.
Hence, calculations of the atomic opacity suitable to calculate the light curve from an early time are necessary.

In this paper, we perform the first systematic atomic opacity calculation for all the heavy $r$-process elements including lanthanides (La - Ra, $Z = 57 - 88$)
ionized to the states V - XI. This work, together with \ct{Smaranikab20, Smaranikab22},
provides the atomic opacity suitable for early time for all the $r$-process elements Ca - Ra ($Z = 20 -88$).
We show our new atomic and opacity calculations in \ar{sec:op}. Using the new opacity dataset, we study the early kilonova emission at $t < 1$ day.
The radiative transfer simulations performed for such purpose are discussed in \ar{sec:lc}.
Since now we have the complete atomic opacity to calculate the light curve for all possible compositions,
we assess the applicability of the gray approximations used in many previous studies for modelling the early kilonova (e.g., \cta{Villar17}) in \ar{sec:disc}.
Finally, we provide our conclusions in \ar{sec:conc}. Note that AB magnitude system is adopted throughout the article.


\begin{figure*}[t]
\begin{tabular}{c}

\begin{minipage}{0.5\hsize}
\begin{center}

\includegraphics[width=\linewidth]{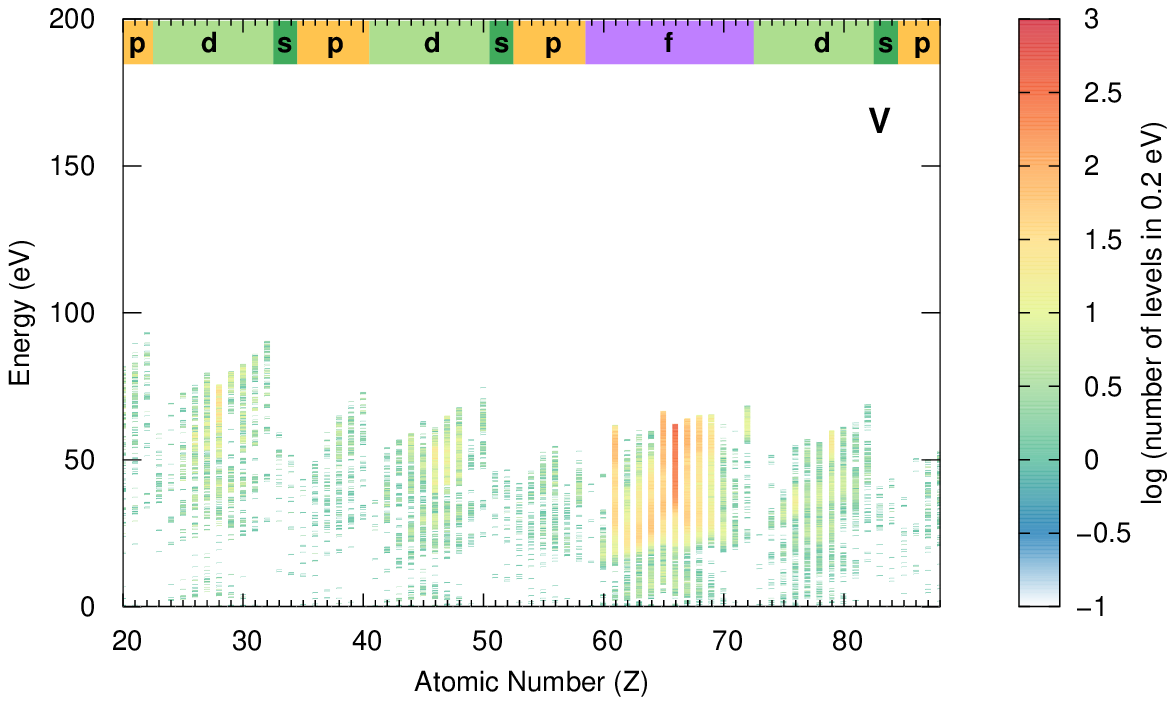}
\end{center}
\end{minipage}

\begin{minipage}{0.5\hsize}
\begin{center}

\includegraphics[width=\linewidth]{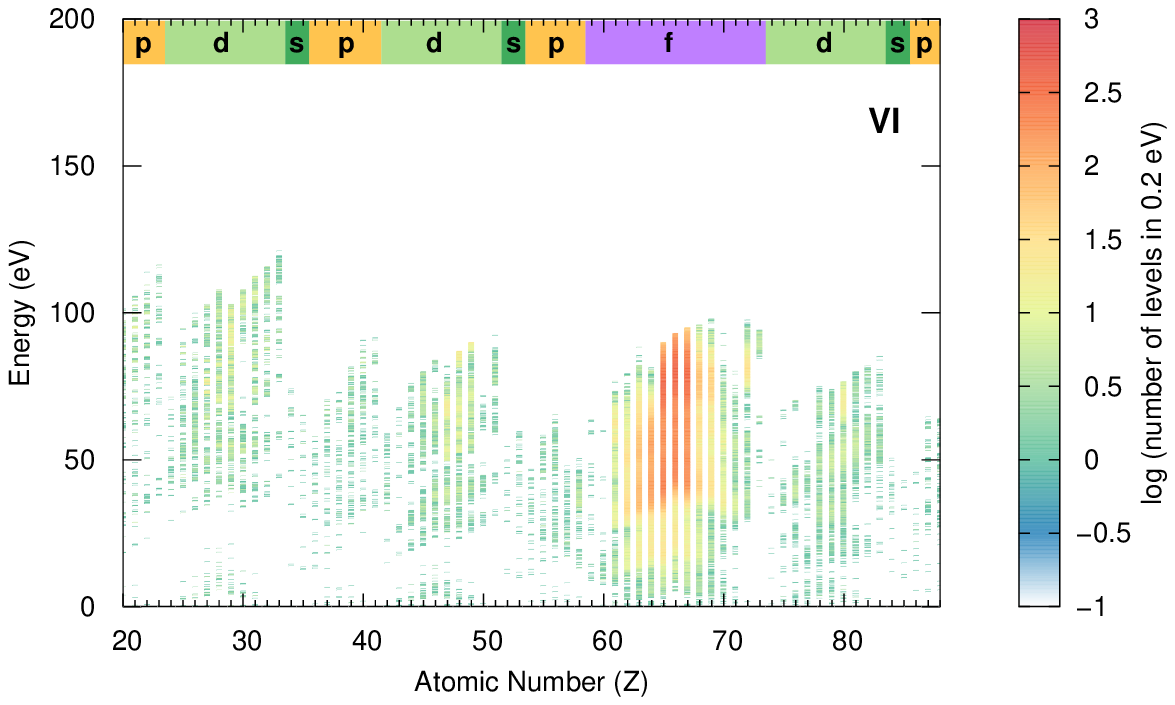}
\end{center}
\end{minipage}

\end{tabular}
\caption{The energy level density distribution for all elements with ionization V (left) and VI (right).
Only the energy levels below the ionization threshold are included.
The colored boxes on the top indicate the valence shells in different ions.}

\label{fig:elev_v_vi}
\end{figure*}
                
\begin{figure*}[t]
\begin{tabular}{c}

\begin{minipage}{0.5\hsize}
\begin{center}
\includegraphics[width=\linewidth]{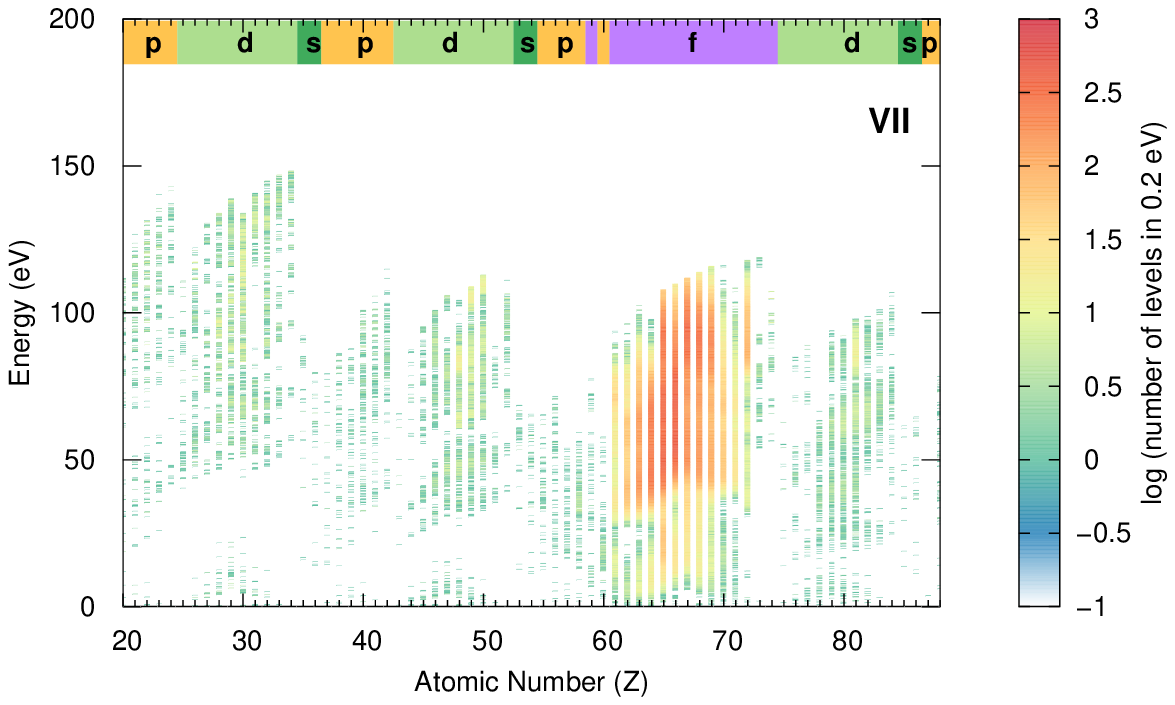}
\end{center}
\end{minipage}

\begin{minipage}{0.5\hsize}
\begin{center}
\includegraphics[width=\linewidth]{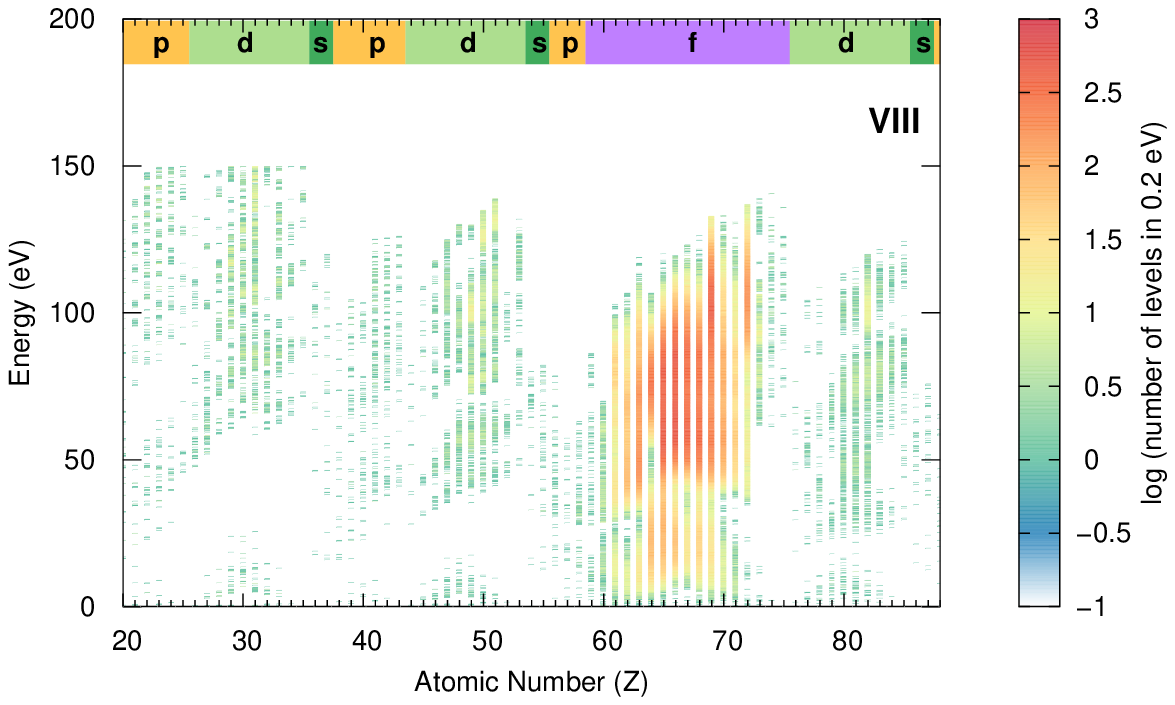}
\end{center}
\end{minipage}

\\

\begin{minipage}{0.5\hsize}
\begin{center}
\includegraphics[width=\linewidth]{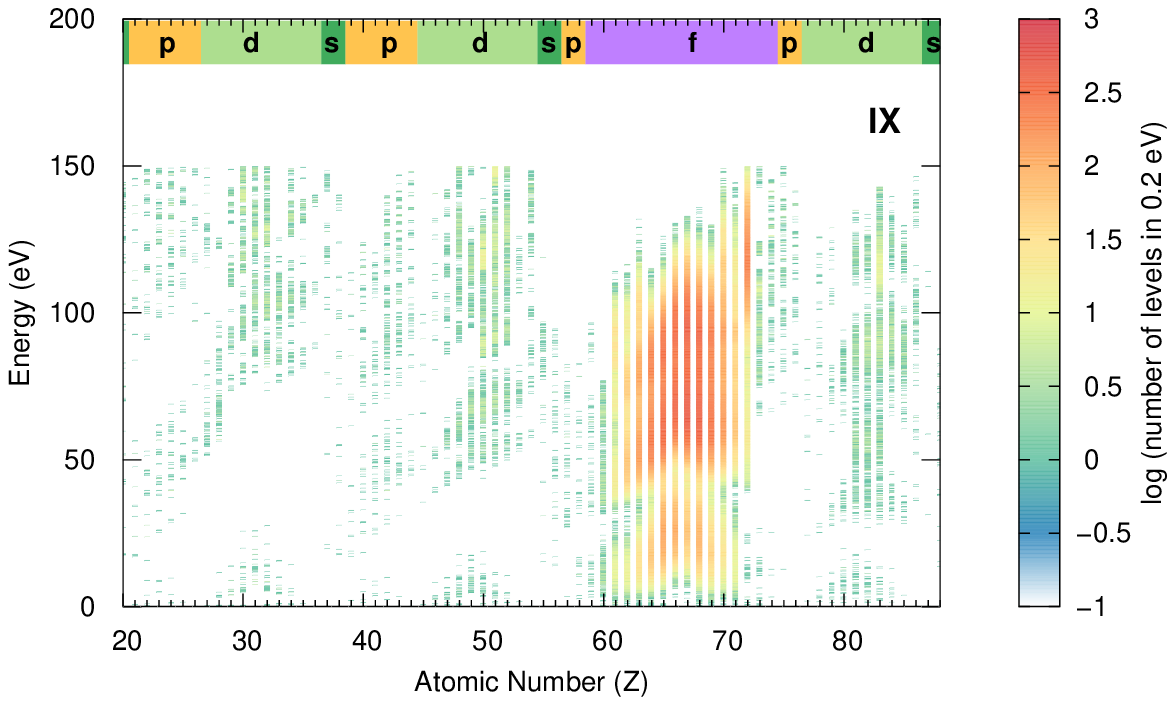}
\end{center}
\end{minipage}

\begin{minipage}{0.5\hsize}
\begin{center}
\includegraphics[width=\linewidth]{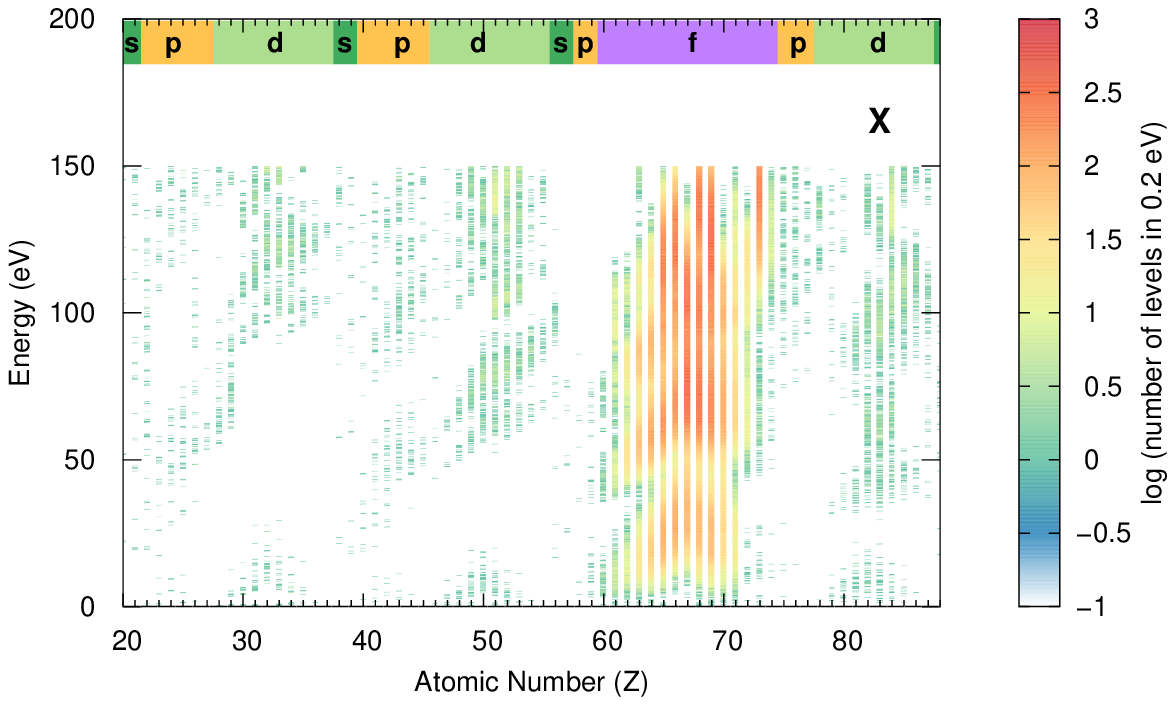}
\end{center}
\end{minipage}

\\

\begin{minipage}{0.5\hsize}
\begin{center}
\includegraphics[width=\linewidth]{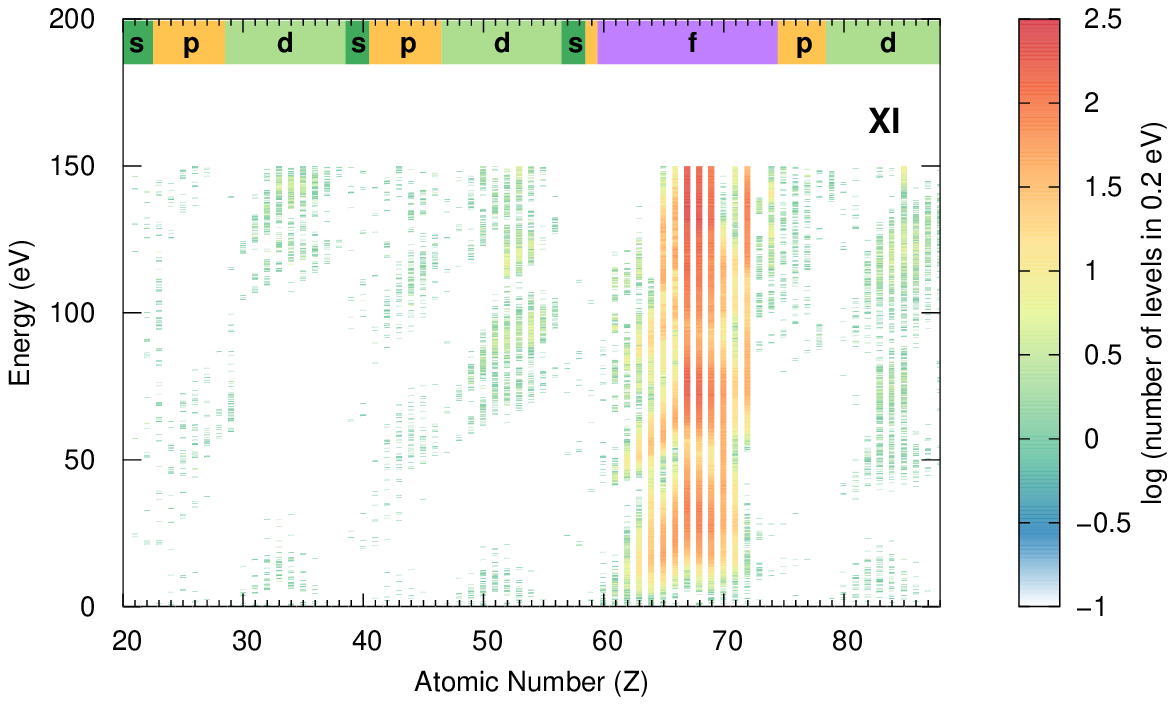}
\end{center}
\end{minipage}

\end{tabular}
\caption{The same as the \ar{fig:elev_v_vi} but for ionization VII - XI.}
\label{fig:elev_vii_xi}
\end{figure*}

\begin{figure*}[t]
\centering
\includegraphics[width=0.5\linewidth]{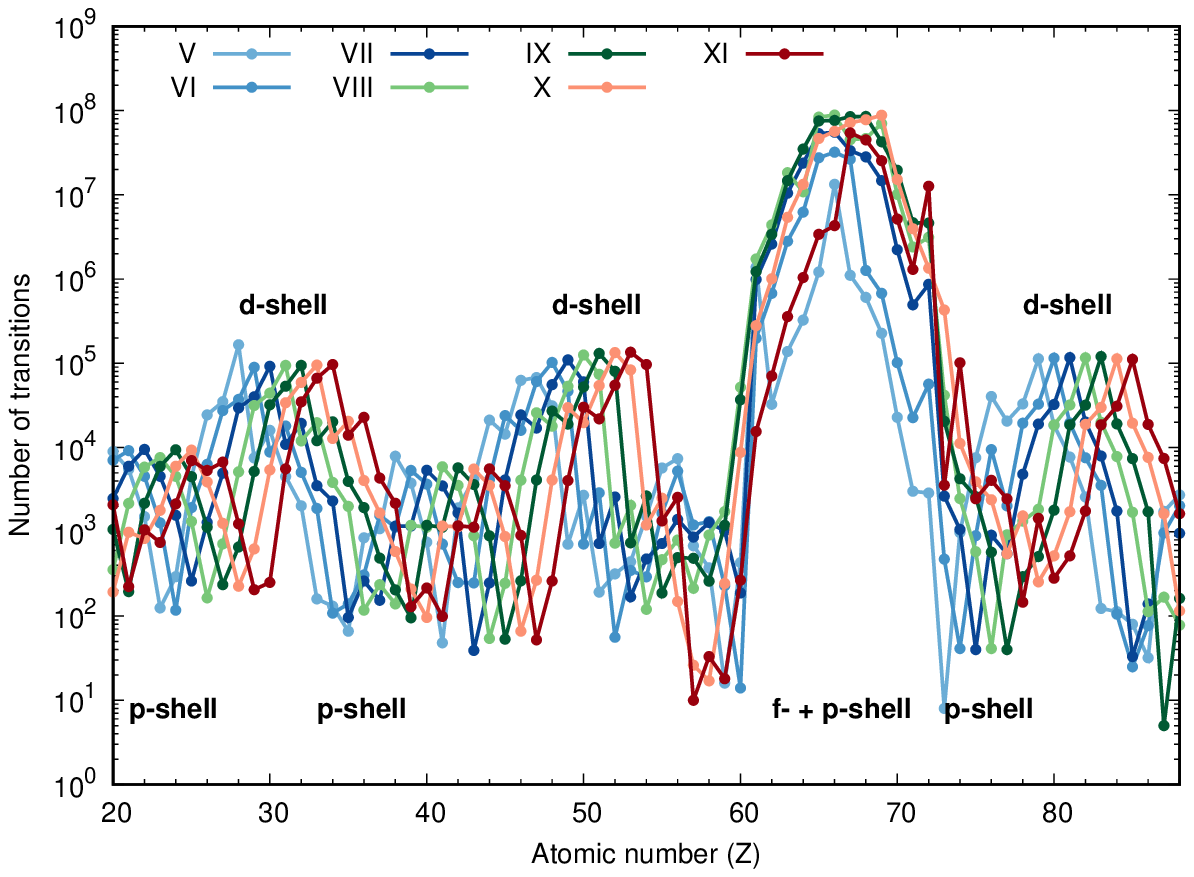} 
\caption{Number of transitions for the ionization V to XI for all the elements from Ca ($Z = 20$) to Ra ($Z = 88$).}
\label{fig:tranZ}
\end{figure*}

\section{Opacity for neutron star merger}\label{sec:op}
In this section, we discuss the bound-bound opacity calculation for the neutron star merger ejecta.
In the case of supernovae and neutron star mergers, the matter is expanding with a high velocity and a high velocity gradient.
In this case, the photons are continuously redshifted in the comoving frame and the redshifted photons progressively interact with lines.
Note that the lines are not infinitely sharp, but are broadened due to the different mechanisms
(mainly due to the thermal motion in neutron star merger ejecta). If the lines do not overlap with each other,
i.e., if the interaction of the photons with different lines are independent of each other, 
then the average contribution from multiple lines within a chosen wavelength bin is defined to be the opacity \citep{Karp77, Eastman93}.
This is called the expansion opacity \ctp{Sobolev60}, calculated as:
\begin{equation}\label{eqn:kexp}
    \kappa_{\rm{exp}}(\lambda) = \frac{1}{\rho ct}\sum_{l}\frac{\lambda_{l}}{\Delta \lambda}(1 -e^{-\tau_{l}}),
\end{equation}
where $\lambda_{l}$ is the transition wavelength in the wavelength interval of $\Delta \lambda$
and $\tau_{l}$ is the Sobolev optical depth at the transition wavelength, calculated as
\begin{equation}\label{eqn:tau}
    \tau_{l} = \frac{\pi e^{2}}{m_{\rm{e}} c} n_{l}\lambda_{l}f_{l}t.
\end{equation}
Here $n_{l}$ is the number density of the lower level of the transition, and $f_{l}$ is the strength of transition.

It is evident from \ar{eqn:kexp} and \ar{eqn:tau} that the calculation of the expansion opacity requires the atomic data
(e.g., energy levels, transition wavelengths, strength of transitions).
Hence, we first perform the atomic structure calculation for the elements La - Ra ($Z = 57 - 88$)
with the ionization states V - XI as discussed in the next section.

Note that the opacity is calculated by assuming the local thermodynamic equilibrium (LTE).
Hence, to calculate the ionization fraction, we solve the Saha ionization equation 
and we determine the population of the excited levels via Boltzmann statistics.
The assumption of LTE is invalid if the non-thermal processes significantly affect the ionization and excitation structure.
However, \ct{Smaranikab22} estimate that at the early times ($t < 1$ day),
the effect of the non-thermal processes on the ionization/excitation is not significant.
Hence, we adopt LTE assumption in this paper.

\subsection{Atomic structure calculations} \label{sec:atom}

We perform atomic structure calculations by using the Hebrew University Lawrence Livermore Atomic Code (HULLAC, \cta{bar-shalom01}).
HULLAC uses zeroth-order solutions of single particle Dirac equations with a parametric central field potential as a basis for perturbation theories.
The central field is parametrized to include the combined effect of the nuclear field and the spherically averaged electron-electron interaction.
The central field is optimized so that the first order configuration averaged energies of the ground and low-lying excited states are minimized.
The zeroth-order wavefunctions are calculated by solving the single electron Dirac equation.
Many-electron wavefunction is constructed from the anti-symmetrized products of the orbitals in any chosen coupling scheme ($j$-$j$ coupling in this case).
Breit interaction and quantum electrodynamic correction are also taken into account.
More details of the HULLAC calculations for heavy elements can be found in the previous studies (e.g., \cta{Tanaka18, Rynkun21, Smaranikab22}).

Note that HULLAC uses the parametric central field potential constructed from the electron distribution of the slater type orbital,
for which the ground state configuration of the next ion state is used.
Hence, the knowledge of the ground configurations are required to perform atomic calculations.
However, for highly ionized atoms, the ground configurations are not well established for several elements.
This problem is especially severe in the case of highly ionized lanthanides ($Z = 57 - 71$).
For example, the ground states for the highly ionized lanthanides provided by the NIST atomic spectra database \ctp{NIST20}
are based on approximated and simplified theoretical calculations \ctp{Carlson70, Rodrigues04, Sugar75a, Martin78}.
Use of such ground states can be a source of uncertainty in the atomic opacity.
Hence, it is important to derive the correct ground configurations to calculate the opacity.

We estimate the ground configurations of the highly ionized lanthanides (V - XI) by using the method originally devised by \citet{Smaranikab22}
to derive the ground configurations of the three lanthanides (Nd, $Z = 60$; Sm, $Z = 62$; Eu, $Z = 63$) ionized up to XI.
We extend the calculation to all the lanthanides.
We perform several atomic structure calculations by systematically varying the central potential.
For designing the central potential, (1) we consider different electron distribution in the $4f$- and $5p$-shells
(the two open shells in the highly ionized lanthanides, \cta{NIST20}) and (2) optimize the central potential with different sets of configurations.
For each case, we identify the ground configuration as the one that generates the configuration state function (CSF)
with the largest mixing coefficient for the lowest energy level.
If the same configuration is identified as the ground state for all the different calculations, we take that as the ground configuration of that ion.
More details on this method can be found in \citet{Smaranikab22}.

Note that our atomic calculation is performed by solving the relativistic Dirac equation
for the individual lanthanide ions unlike the previous simplified theoretical calculations \ctp{Carlson70, Rodrigues04, Sugar75a, Martin78}.
Hence, we use the derived ground configurations for the highly ionized lanthanides for the atomic structure calculations.

We show all the configurations used in our work in \ar{tab:configtbl}, where the first configuration always represents the ground one.
All the configurations used for optimization of the central potential are marked in bold in \ar{tab:configtbl}. 
Note that for the post-lanthanides (elements with atomic number $Z$ = 72 $-$ 88),
we use the ground states provided in the NIST \ctp{NIST20}, following the previous works (e.g., \cta{kasen13, Tanaka20a, Smaranikab20}).

\begin{figure*}[t]
\begin{tabular}{c}

\begin{minipage}{0.5\hsize}
\begin{center}
\includegraphics[width=\linewidth]{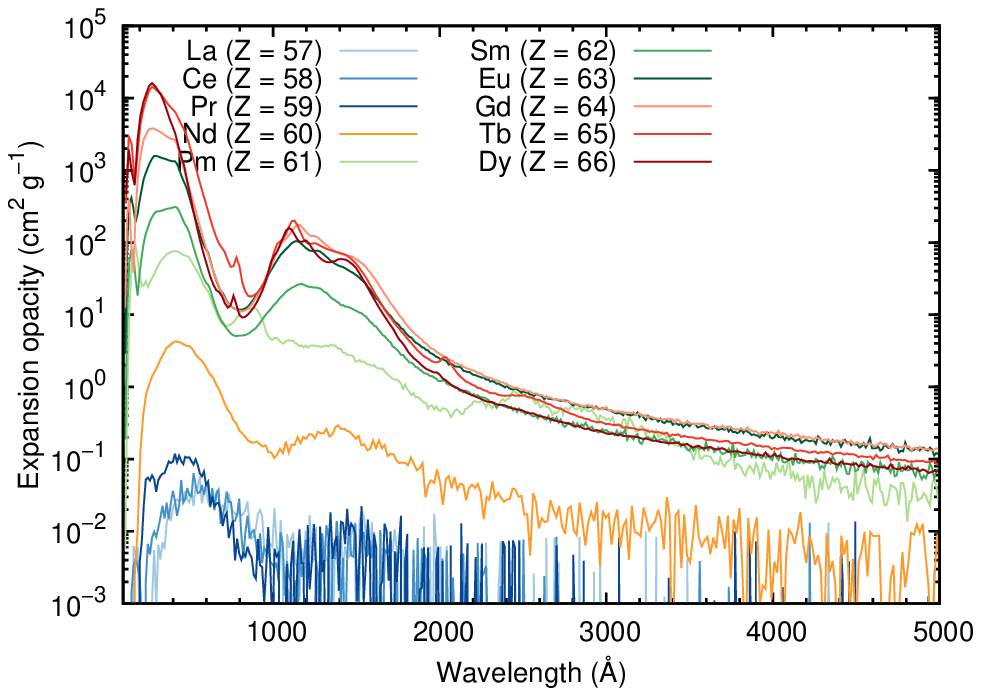}
\end{center}
\end{minipage}

\begin{minipage}{0.5\hsize}
\begin{center}
\includegraphics[width=\linewidth]{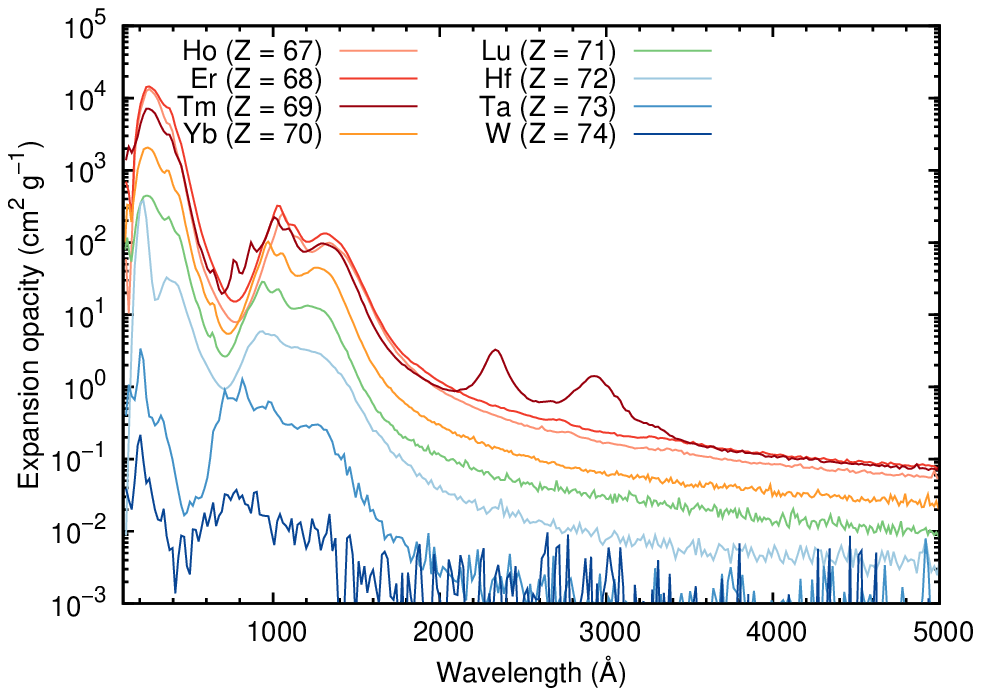}
\end{center}
\end{minipage}

 \\
\begin{minipage}{0.5\hsize}
\begin{center}
\includegraphics[width=\linewidth]{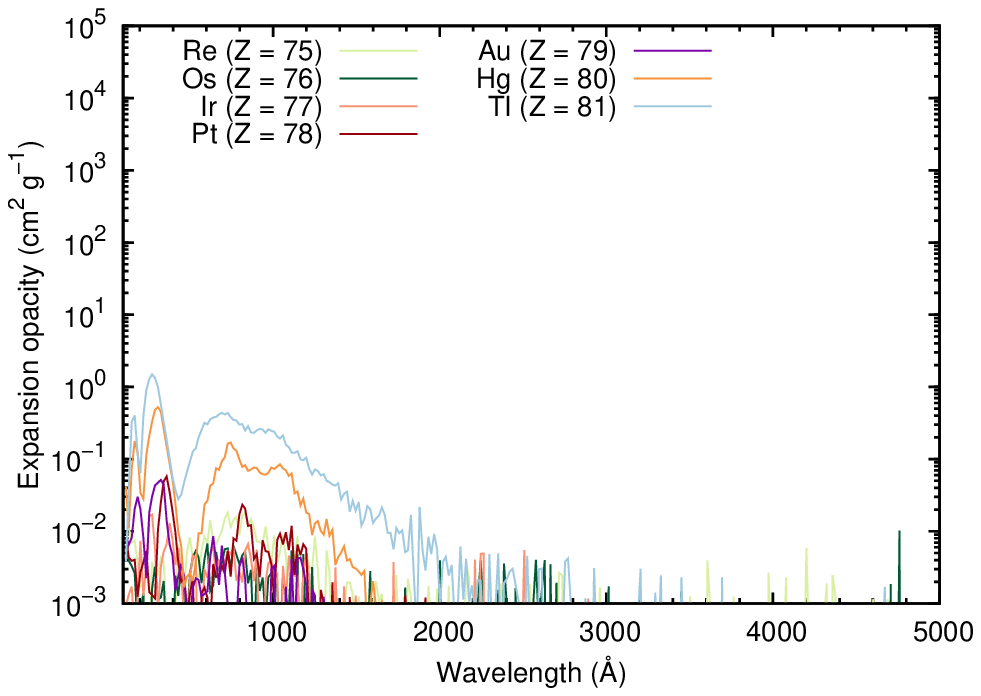}      
\end{center}
\end{minipage}

\begin{minipage}{0.5\hsize}
\begin{center}
\includegraphics[width=\linewidth]{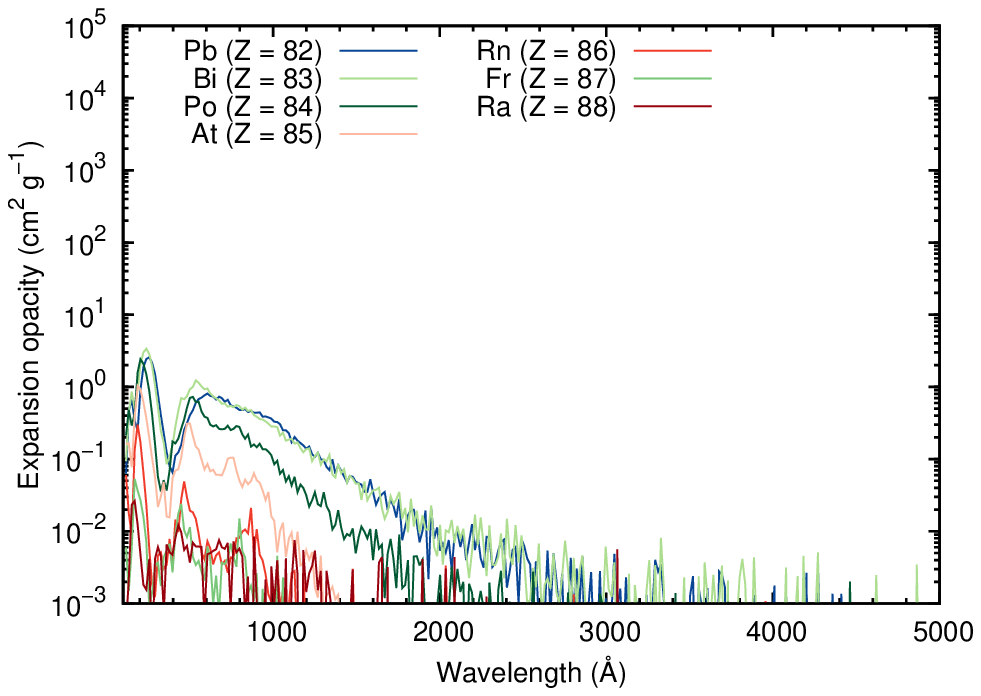}      
\end{center}
\end{minipage}

\end{tabular}
\caption{The expansion opacity as a function of wavelength at $T\,=\,70,000$ K, $\rho \,=\, 10^{-10}\, \rm g\,cm^{-3}$,
  and at $t = 0.1$ days for the different elements.}
\label{fig:expwav_fdps}

\end{figure*}


\begin{figure*}[t]
\begin{tabular}{c}

\begin{minipage}{0.5\hsize}
\begin{center}
\includegraphics[width=\linewidth]{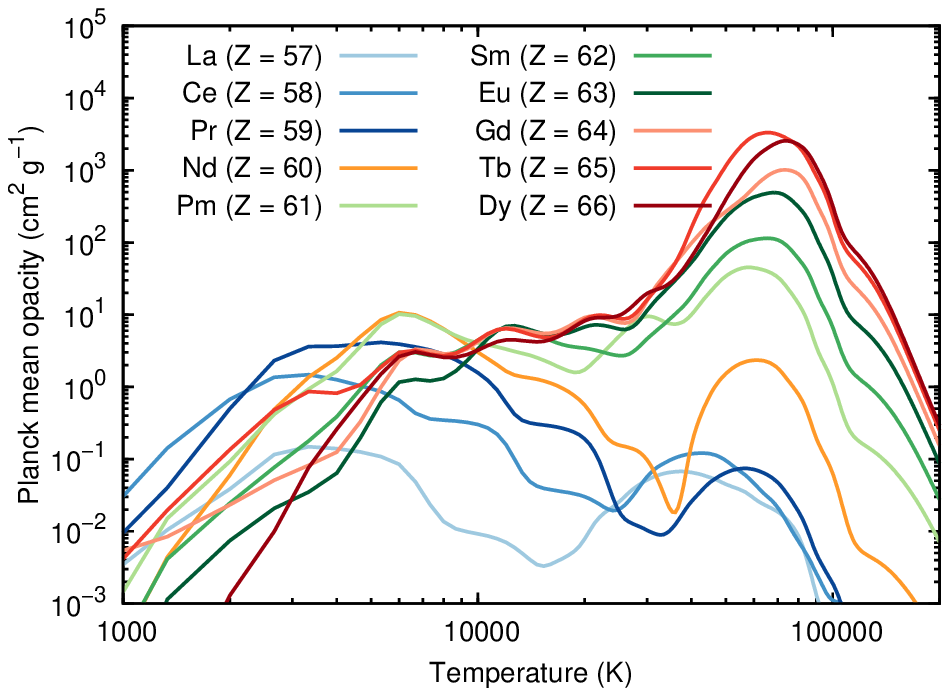}
\end{center}
\end{minipage}

\begin{minipage}{0.5\hsize}
\begin{center}
\includegraphics[width=\linewidth]{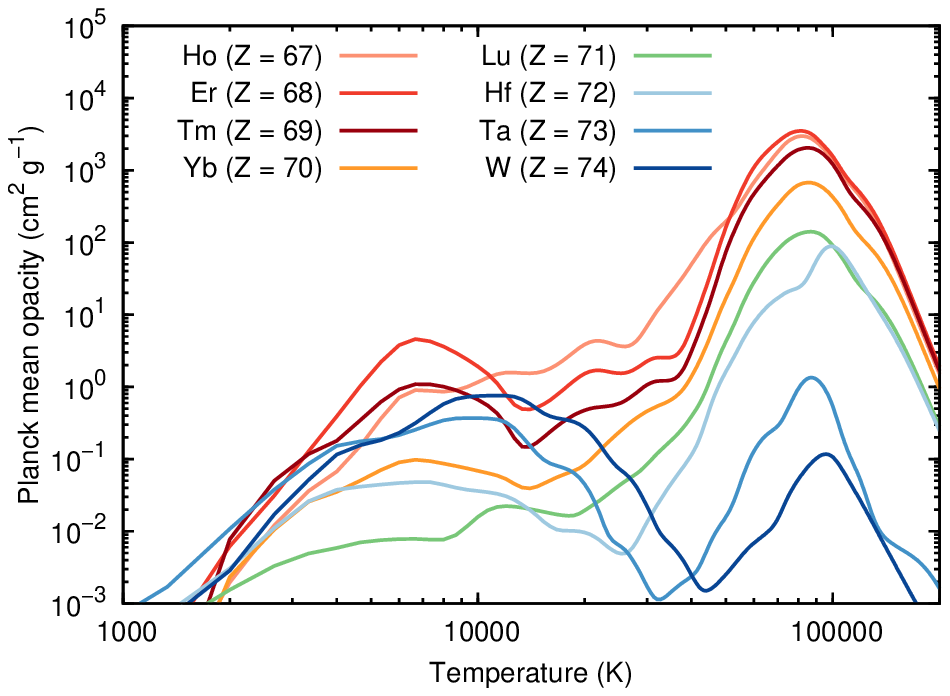}
\end{center}
\end{minipage}

\\
\begin{minipage}{0.5\hsize}
\begin{center}
\includegraphics[width=\linewidth]{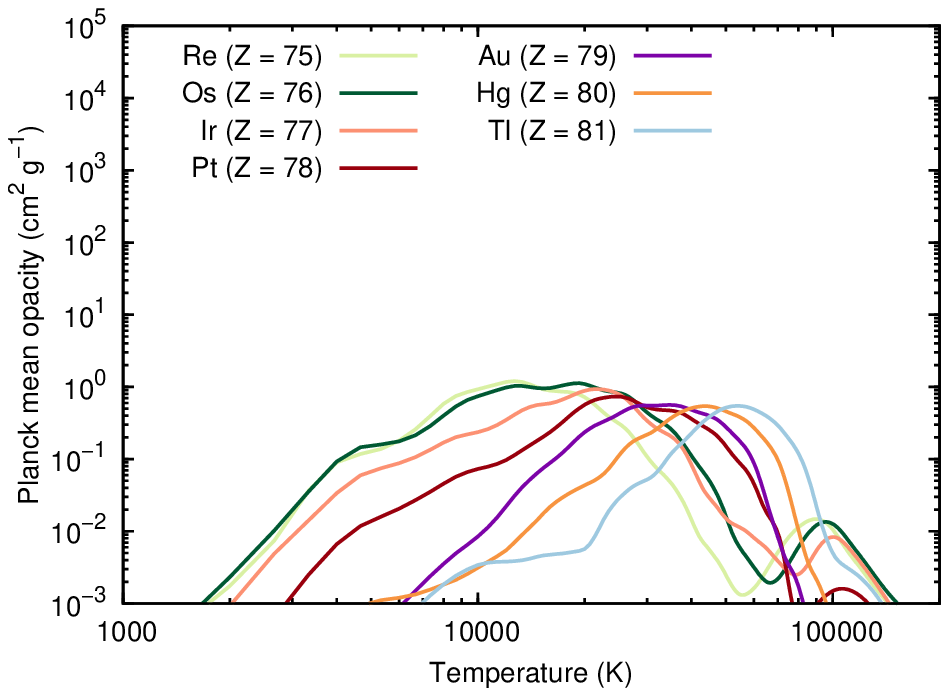}      
\end{center}
\end{minipage}

\begin{minipage}{0.5\hsize}
\begin{center}
\includegraphics[width=\linewidth]{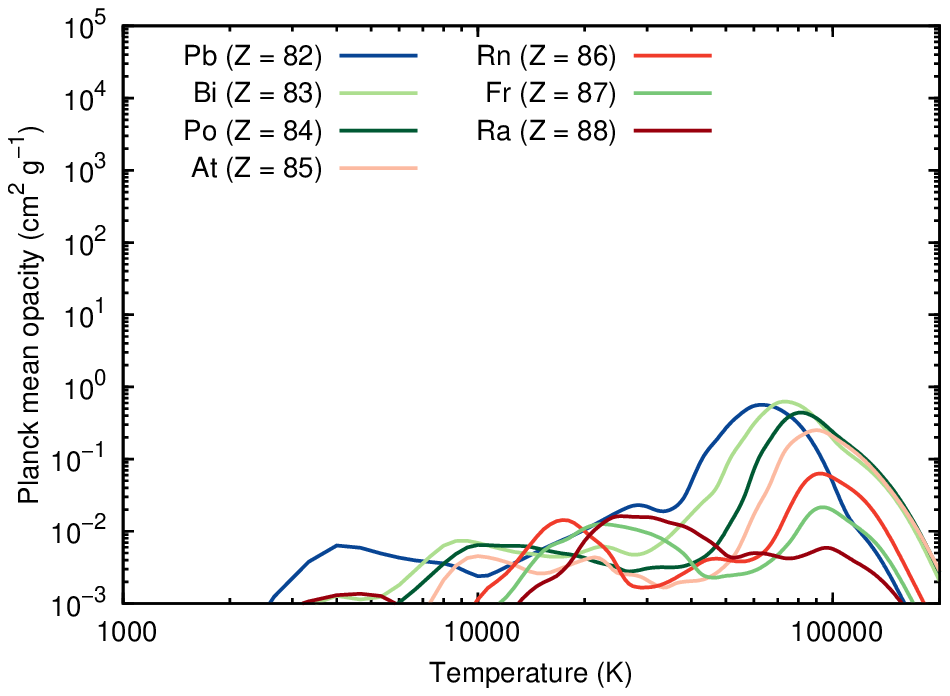}      
\end{center}
\end{minipage}

\end{tabular}
\caption{The Planck mean opacity as a function of temperature at $\rho\, =\, 10^{-10} \,\rm g\,cm^{-3}$ and $t = 0.1$ days
for different elements.} 
  
\label{fig:expT_fdps}

\end{figure*}

\subsection{Energy levels and transitions}\label{sec:atom_res}

\ar{fig:elev_v_vi} and \ar{fig:elev_vii_xi} show the energy level distribution obtained from our atomic structure calculations.
We also include the lighter $r$-process elements ($Z = 20 - 56$, \cta{Smaranikab20})
and the three lanthanides (Nd, Sm, Eu) already discussed in \ct{Smaranikab22} for comparison.
Note that the number of energy levels below the ionization threshold are shown 
since these are the only levels that are relevant for the bound-bound opacity.
The color scale represents the level density in the 0.2 eV energy bin.
The colored boxes on the top of the figure indicate the valence electron shell of the individual ions.

Note that the valence electron shells in highly ionized states are different than the valence shells in neutral atoms.
For example, Pb ($Z = 82)$, a $p$-group element at neutral state, has $d$-shell as the outermost orbital in ionization IX (\ar{tab:configtbl}).
Again, Hf ($Z = 73$), which is a $d$-shell element at neutral state, has $f$-shell as valence shell at higher ionization ($>$ V, \ar{tab:configtbl}).
Depending on the valence shell, the energy level distribution of a particular ion might show completely different trend than its neutral or lower ionization state.

The energy level density in the highly ionized elements is the maximum when the $f-$shell electrons are present.
For instance, the energy level density is extremely high for the elements Pr - Ta ($Z = 59 -73$),
which have open $f$-shell as the valence shell for ionization states of V - XI.
This is due to the large number of available states when the open $f-$shell is present.
This behavior for the elements with $f-$shell is well known for the low ionization cases
($Z = 57 - 71$, e.g., see \cta{kasen13, Tanaka13, Fontes15, Fontes20, Fontes22, Wollaeger17, Tanaka18, Tanaka20a}).
However, the energy level density for the elements with open $f-$shell at the higher ionization are even higher.
This is because at higher ionization, the elements with open $f$-shell possess the open $p$-shell in addition (\ar{tab:configtbl}),
increasing the total number of the available states (or complexity) for the electrons.
The high density of the energy levels for the highly ionized lanthanides is also observed by
\ct{Smaranikab22} for Nd ($Z = 60$), Sm ($Z = 62$), and Eu ($Z = 63$).
Now the overall trend for highly ionized lanthanides is clearer with the extention of the calculations to all the lanthanides.

For the other elements, the energy level density is the maximum for the elements with open $d$-shell,
followed by the elements with open $p$- and $s$-shell (\ar{fig:elev_v_vi}, \ar{fig:elev_vii_xi}).
For example, at ionization IX, the density of energy levels appear in the decreasing order for the elements which possess open $d$-shell (Ir - Rn, $Z = 77 - 86$),
open $p$-shell without any $f$-shell (W - Os, $Z = 74 - 76$), and open $s$-shell (Fr - Ra, $Z = 87 - 88$; \ar{fig:elev_vii_xi}).
Note that the similar trends are followed for the lighter $r$-process elements with open $d$-, $p$-, and $s$-shell \ctp{Smaranikab20}.

The energy level density for the elements with the same open shell becomes the maximum when the open shells are half-filled.
This is because the half-filled shells have the maximum complexity.
For example, for the elements with open $d$-shell at ionization IX (Ir - Rn, $Z = 77 - 86$), 
the maximum energy level densities appear around Tl - Po ($Z = 81 - 83$), where the open $d$-shell is nearly half-filled (\ar{tab:configtbl}).
Similar trends are observed for the elements with the other open shells, e.g., the elements with $p$-shell and $s$-shell across different ionizations.

In case of the elements with open $f$-shell at higher ionization, open $p$-shell is also present.
The number of electrons in the two open shells affects the energy level density distributions in different ways:
(1) for a particular ionization, the number of electrons in the $f-$shell determines the elements which have the most dense energy level density structure.
For instance, the energy level density is the maximum for the elements Tb to Tm ($Z = 65 - 69$, see \ar{fig:elev_v_vi}, \ar{fig:elev_vii_xi}),
which have the nearly half-filled $f$-shell (\ar{tab:configtbl}) across different ionizations;
(2) for a particular element, the number of electrons in the $p$-shell decides the ionizations at which the energy level density reaches its maximum.
As an example, consider Eu, for which the energy level density distribution is the maximum around the ionization VIII - X (\ar{fig:elev_vii_xi}),
where the $p$-shell is nearly half-filled (\ar{tab:configtbl}).
If both the shells are half-filled, as that in the case for the elements Tb - Tm ($Z = 65 - 69$) at the ionization states of $\sim$ VIII - X,
then they show the highest energy level densities among all the different lanthanide ions (\ar{fig:elev_v_vi}, \ar{fig:elev_vii_xi}).

\ar{fig:tranZ} shows the total number of transitions for different elements ($N_{\rm line}$ in \ar{tab:configtbl}) in various ionization states (V - XI).
The labels provide the valence shells for the element with highest number of transitions at a particular ionizations
(note that we do not specify the ions with $s$-shell explicitly, nevertheless, these can be found in \ar{tab:configtbl}).
The lighter $r$-process elements ($Z \le 56$, \cta{Smaranikab20}) and the three lanthanides (Nd, Sm, Eu, \cta{Smaranikab22}) are also shown for comparison.

The trend in the total number of transitions reflects the number density of the energy levels (\ar{fig:elev_v_vi}, \ar{fig:elev_vii_xi}).
For instance, at any particular ionization, the number of transitions is consecutively lower for the elements with open $f$- and $p$-shell,
followed by the elements with open $d$-, $p$-, and $s$-shell (\ar{fig:tranZ}).
Moreover, the transitions for the elements with the same open shells becomes the maximum when the open shells are half-filled.
This is due to the fact that the ions with half-filled shell show the largest number of energy levels and transitions.

For a few of highly ionized elements (VII - XI) with open $f$-shell and $p$-shell, 
the total number of transitions involving all the energy levels ($N_{\rm line}^{*}$ in \ar{tab:configtbl})
are higher than the transitions involving the energy levels only below the ionization threshold
($N_{\rm line}$ in \ar{tab:configtbl}, e.g., see Tb VIII in \ar{tab:configtbl}). This can be understood in the following way.
Firstly, most of these ions have high complexity due to the presence of open $f$-shell and $p$-shell, hence,
the total number of energy levels are extremely high.
Also, for these ions, the energy levels are pushed upward due to several effects, such as,
the higher ionization degree, higher spin-orbit interaction \ctp{Cowan81, Tanaka20a}.
Combining these two effects, the transitions involving the higher lying energy levels beyond the threshold energy increases significantly.



\begin{figure*}[t]
\begin{tabular}{c}

\begin{minipage}{0.5\hsize}
\begin{center}
\includegraphics[width=\linewidth]{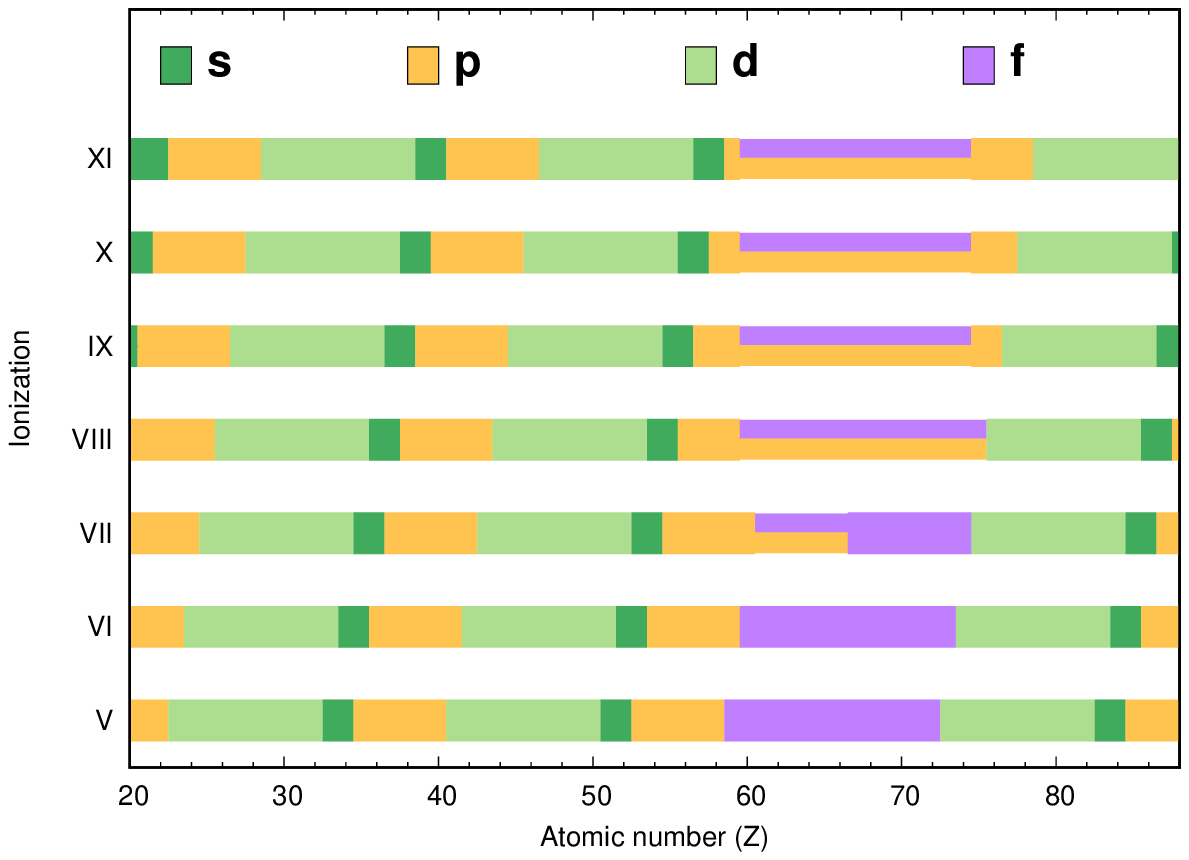}
\end{center}
\end{minipage}

\begin{minipage}{0.5\hsize}
\begin{center}
\includegraphics[width=\linewidth]{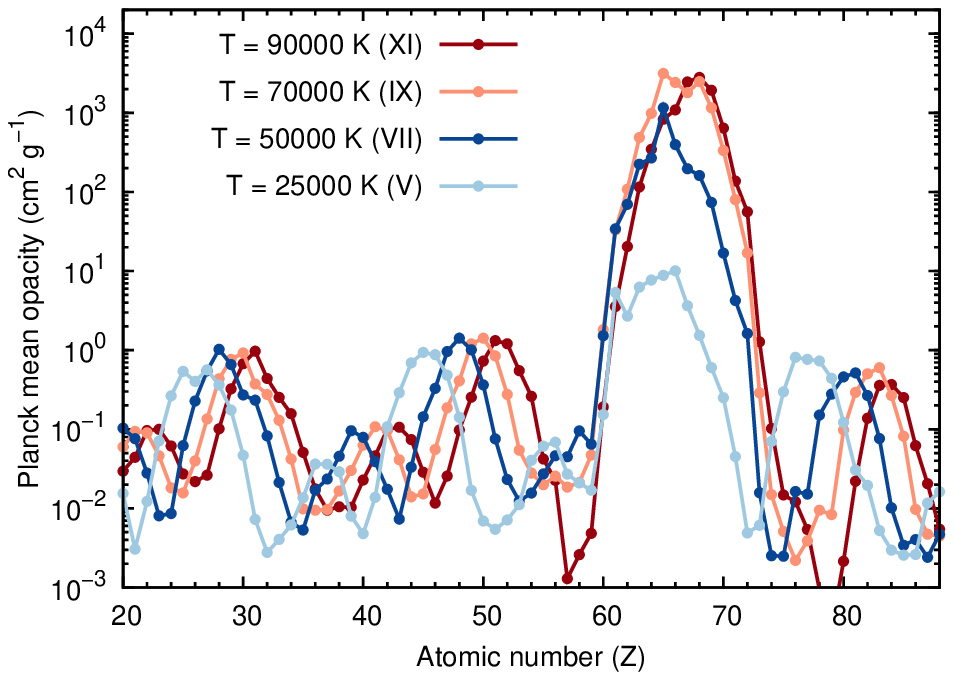}      
\end{center}
\end{minipage}

\end{tabular}
\caption{\textbf{Left:} The change of the valence shell with ionization for different elements.
Note that $s$-, $p$-, $d$-, and $f$-shells are represented by the colors bottle green, yellow, light green, and purple, respectively.
\textbf{Right:} The mean opacity for different elements. Diffferent colors represent different temperatures.}
  
\label{fig:ion_zT}

\end{figure*}


\subsection{Bound-bound opacity}\label{subsec:elem_op}
Using the newly constructed atomic data, we calculate the expansion opacity for all the elements La - Ra ($Z = 57 - 88$) ionized to the states of V - XI.
We assume a single element ejecta with the density $\rho = 10^{-10}\,\rm g\,cm^{-3}$,
which represents the ejecta condition at $t \sim$ 0.1 days for a AT2017gfo-like neutron star merger ejecta with an ejecta mass of $M_{\rm ej}\sim 0.05M_{\odot}$
\ctp{Kasliwal17, kasen17, Waxman18, Smaranikab20}, moving with an average velocity of $v_{\rm avg} \sim 0.1c$ (e.g., \cta{Tanaka17}).
The single-element ejecta are not realistic as the ejecta always contain the mixture of elements.
However, we discuss the single-element opacity to understand the effect of the newly calculated atomic data on the opacity.

In the early time, the outermost layer opacity determines the shape of the light curves \ctp{Smaranikab20, Smaranikab22}.
The temperature of the outermost layer for the ejecta of the AT2017gfo-like neutron star merger ejecta moving with a velocity of
$v \sim 0.2c$ (e.g., \cta{Tanaka17}) is $T\sim 70,000 $ K \ctp{Smaranikab20}.
Since our main aim is to provide the light curve for the early kilonova, we choose this temperature to discuss the expansion opacity.
Around the temperature of $T = 70,000$ K, the elements become IX ionized.

The expansion opacities for different elements at a temperature of $T = 70,000$ K are shown in \ar{fig:expwav_fdps}.
The expansion opacities show a strong wavelength dependence, with a higher value at the shorter wavelengths ($\lambda < 1000 \, \rm \AA$).
This is because of the higher number of transitions at shorter wavelengths.



The expansion opacities for the elements with different shells vary widely. 
For example, the opacities for the elements with open $f$- and $p$-shell ($Z = 59 - 73$ at ionization IX,
relevant at temperature $T \sim 70,000$ K) are extremely high, reaching $\kappa_{\rm exp} = 10^4\, \rm cm^2\,g^{-1}$ at $\lambda \le 1000\, \rm \AA$, 
followed by the elements with open $d$-shell ($Z = 77 - 86$ at ionizations IX),
open $p$-shell without $f$-shell ($Z \sim 74 - 76$ at ionizations IX), and with open $s$-shell ($Z \sim 87 - 88$ at ionizations IX),
with the opacity reaching $\kappa_{\rm exp} = 1.0\, \rm cm^2\,g^{-1}$, $0.1\, \rm cm^2\,g^{-1}$, and $0.01\, \rm cm^2\,g^{-1}$, respectively,
at the same wavelengths.

The wide range of the expansion opacities across the elements with different valence shells stems from the differences
in the energy level density (and correspondingly, the number of transitions) in the presence of the different open shells (see \ar{sec:atom_res}).
Note that this is different from the low ionized case (e.g., \cta{Tanaka20a}), where the density of those energy levels determines the opacity because
only the low-lying energy levels are populated at lower ionizations (i.e., at low temperature).
Hence, the number density of the low-lying energy levels decides the opacity. On the other hand, in the highly ionized state (i.e., at a higher temperature),
even the higher energy levels can be populated. Hence, the total energy level density is more important to determine the opacity.

We discuss the limitation of the opacity calculation before proceeding further.
For the expansion opacity in the neutron star mergers, the contribution of multiple lines are taken into account
under the assumption that the interactions between the photons and lines are independent of each other.
This assumption requires the photons to travel freely between the two different interactions with different lines.
Such assumption works fine as long as the resonance zones or the line profiles do not overlap each other. 
However, if the line profiles start to overlap with each other, assuming the expansion opacity underestimates the opacity.
This is because, in such case, taking the expansion opacity overestimates the photon mean free path by assuming independent interactions.

To indicate when the line profiles start overlapping, a critical opacity is defined as
$\kappa_{\rm{crit}} \sim 3\, \rho_{-10}^{-1} \,v_{\rm th,\, 4}^{-1} \, t_{0.1}^{-1} \, \rm {cm^{2}\,g^{-1}}$
\ctp{kasen13, Smaranikab22}, where $\rho_{-10} = \rho/ 10^{-10}\, \rm {g\,cm^{-3}}$,
$v_{\rm th,\, 4}$ is the thermal velocity $v_{\rm th}/4\, \rm{km\,s^{-1}}$, the typical values at $t_{0.1} = t/0.1\,\rm day$.
If the expansion opacity is higher than this critical value, that implies that the line profiles start overlapping.
The expansion opacities for the elements with open $f$-shell or the lanthanides are higher than the critical value at far-UV ($\lambda \le 1000\, \rm \AA$).
Hence, the opacities in the far-UV wavelengths for the lanthanides are mostly underestimated.


\begin{figure*}[t]
\centering
\begin{tabular}{c}
\includegraphics[width=0.5\linewidth]{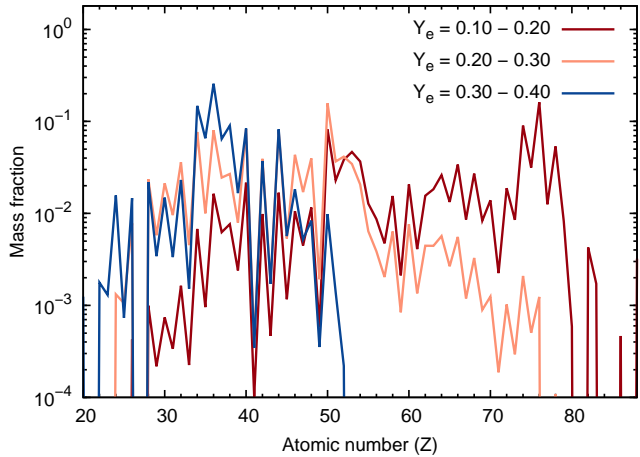}   

\end{tabular}

\caption{The abundance pattern for different electron fraction. The data are taken from \ct{Wanajo14}.}

\label{fig:abun}
\end{figure*}

Keeping the limitation of the opacity calculation in mind,
we calculate the Planck mean opacities to study the temperature dependence of the expansion opacity.
For that purpose, we convolve the expansion opacities with the blackbody function.
Note that the Planck mean opacity is underestimated for the lanthanides at higher temperature since
the blackbody function peaks at far-UV at higher temperature, where the expansion opacity exceeds the critical opacity.

The peaks of the opacity for lanthanides appear at $T \sim 70,000 - 90,000\, \rm K$.
This is because lanthanides become $\ge$ VII ionized around this temperature, where the ions have two open shells ($f$- and $p$-shell, \ar{tab:configtbl}).
Note that for different lanthanides, the ionization of $\ge$ VII appears at nearly same temperature since the potential energy does not vary much between lanthanides.
Moreover, among the lanthanides with ionization $\ge$ VII, the maximum value of the opacity appears for the elements Tb to Tm ($Z = 65 - 69$),
which have two open shells nearly half-filled at such ionizations.

For the elements with other valence shells ($d$-shell, $p$-shell without $f$-shell, and $s$-shell),
the maxima in the opacity appears whenever the valence shells become half-filled. For example, at $T \sim 10^5$ K, Re ($Z = 75$) shows the maxima in opacity. 
This is because at this temperature, Re is ionized to $\sim$ XI, where the valence $p$-shell is nearly half-filled.
The same trend is also observed for the $d$-shell, $p$-shell without $f$-shell, and $s$-shell elements for the lighter $r$-process elements \ctp{Smaranikab20}.

The overall variation in the Planck mean opacities for the elements are visualized in (right panel of \ar{fig:ion_zT}),
which shows the Planck mean opacities at the temperatures, $T = 25000$ K, $T = 50,000$ K, $T = 70,000$ K, and $T = 90,000$ K.
Note that these are the temperatures where the peak of the ionization states of V, VII, IX, and XI appear.
It is clear that the valence shells for different elements change with ionizations (left panel of \ar{fig:ion_zT}), and affects the overall opacity.  
Note that the peak opacity in the elements with the two open shells (open $f$- and $p$-shell) are underestimated due to taking the assumption of the expansion opacity.

We discuss the opacity trends for a single element ejecta with a fixed density.
However, such conditions are not realistic for the neutron star merger ejecta, since the neutron star merger ejecta consist of the mixture of elements.
Also, as the ejecta expand with time, the density and temperature change, causing the change in the opacity.
We discuss the opacity of the mixture of elements for the realistic ejecta conditions in the context of the radiative transfer simulations in the next sections.

\section{Radiative transfer simulations} \label{sec:lc}

\begin{table*}[]
\centering
\caption{Models and the corresponding gray opacities}
\label{tab:model_gray}
\begin{tabular}{cccc}
\hline
\multirow{2}{*}{Models} & \multicolumn{2}{c}{Electron fraction}
& \multirow{2}{*}{\begin{tabular}[c]{@{}c@{}}Suggested gray opacities \\ ($\rm cm^2\,g^{-1}$)\end{tabular}} \\ \cline{2-3}
                             & \begin{tabular}[c]{@{}c@{}}$Y_{\rm e, in}$\\ ($v = 0.05c - 0.2c$,\\ $M_{\rm ej, in} = 0.01M_{\odot}$)\end{tabular}
                             & \begin{tabular}[c]{@{}c@{}}$Y_{\rm e, out}$\\ ($v = 0.2c - 0.33c$, \\ $M_{\rm ej, out} = 0.001M_{\odot}$)\end{tabular}
                             &                                                                                                           \\ \hline
Model 1                            & $0.10 - 0.20$                                                                                 & $-$                                                                                              & $2.0 - 10.0$                                                                                                     \\
Model 2                            & $0.20 - 0.30$                                                                                 & $-$                                                                                              & $1.0 - 5.0$                                                                                                   \\
Model 3                            & $0.30 - 0.40$                                                                                 & $-$                                                                                              & $0.8$                                                                                                  \\ \hline
Model 4                            & $0.30 - 0.40$                                                                                 & $0.10 - 0.20$                                                                                    & $5.0 - 20.0$                                                                                                 \\
Model 5                            & $0.30 - 0.40$                                                                                 & $0.20 - 0.30$                                                                                    & $3.0 - 10.0$                                                                                                   \\
Model 6                            & $0.30 - 0.40$                                                                                 & $0.30 - 0.40$                                                                                    & $0.8 - 1.0$                                                                                                  \\ \hline
\end{tabular}
\end{table*}


\begin{figure*}[t]
\begin{tabular}{c}

\begin{minipage}{0.5\hsize}
\begin{center}
\includegraphics[width=\linewidth]{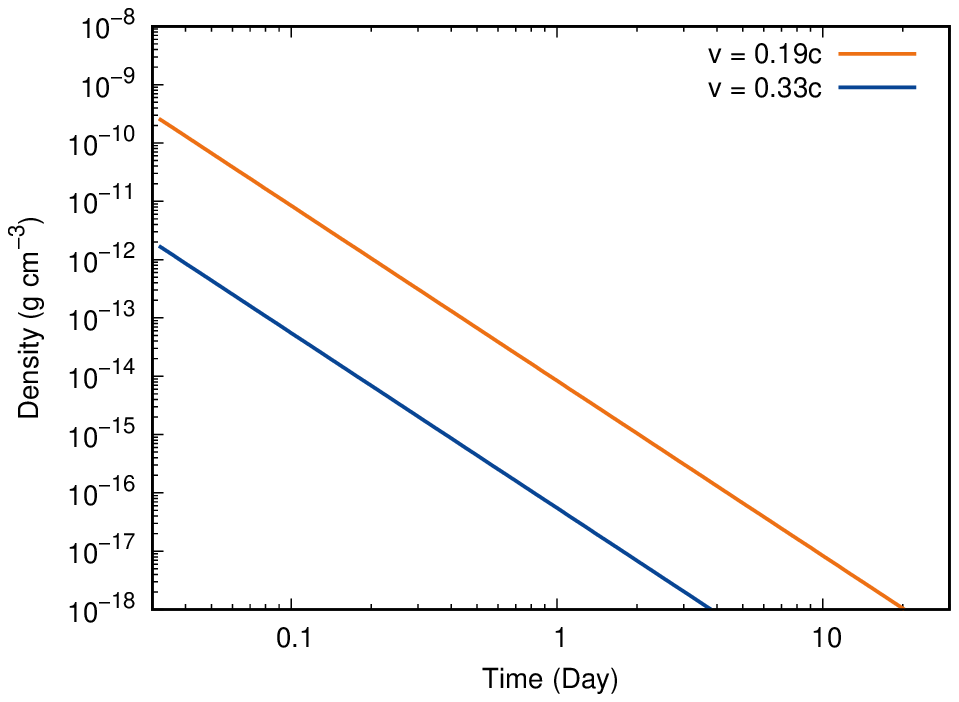}
\end{center}
\end{minipage}

\begin{minipage}{0.5\hsize}
\begin{center}
\includegraphics[width=\linewidth]{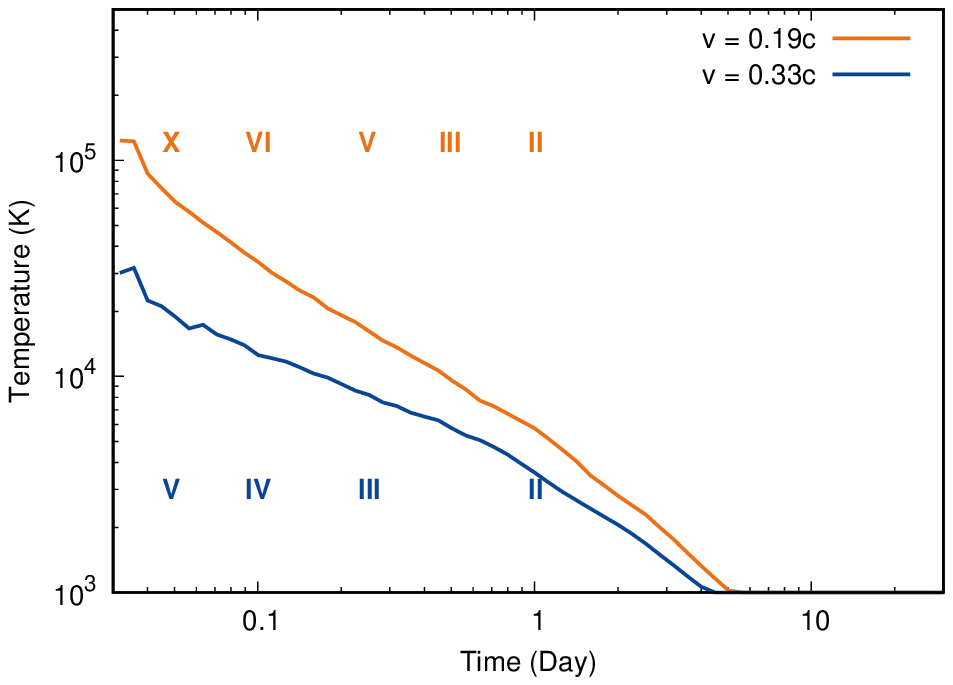}
\end{center}
\end{minipage}
\end{tabular}
\caption{The density (left) and the temperature (right) evolution at the fixed ejecta point,
$v$ = 0.19c and $v$ = 0.33c for model 3 (orange) and model 6 (blue), recpectively. 
In presence of the lanthanides in the ejecta, the light curves are fainter.
Also, the light curves are affected by the thin outer layer in the ejecta.}

\label{fig:rhoT_t}
\end{figure*}

\subsection{Models} \label{sec:lc_model}

In the binary neutron star merger, the elemental distribution dominantly depends on the electron fraction ($Y_{\rm{e}}$), 
i.e., the electron to baryon ratio in the ejecta.
In neutron star merger, masses are ejected in several different channels \ctp{Shibata19}, producing multiple ejecta components.
Different ejecta components have different electron fraction. 
For the dynamical ejecta, the mass ejected towards the polar direction due to the shock between the interface of the neutron stars
have relatively higher electron fraction ($Y_{\rm{e}}>0.25$, e.g., \cta{Bauswein13, Just15, Sekiguchi15, Just22}),
whereas the mass ejected due to the tidal disruption of the neutron stars have relatively lower electron fraction
($Y_{\rm{e}}<0.25$, e.g., \cta{Bauswein13, Just15, Sekiguchi15, Just22}).
Additionally, masses are ejected (more isotropically) from the disk formed around the central remnant in a longer timescale.
This disk wind ejecta have relatively higher electron fraction due to the higher neutrino irradiation
(e.g., \cta{Metzger14, Miller19, Fernandez14, Perego14, Lippuner17, Fujibayashi18, Fujibayashi20a, Fujibayashi20b}).
Differences in the electron fraction introduce variance in the elemental abundance pattern in different ejecta components,
which introduces a viewing angle dependence in the kilonova light curves.

To understand the early kilonova for different compositions, we perform radiative transfer simulations 
for an 1D spherical ejecta model \citep{Metzger10} with a power-law density structure $\rho \propto r^{n}$,
where $n = -3$, moving with the velocity range of $v\,=$ 0.05c to 0.2c. The electron fraction is taken to be homogeneous with the values:
(1) $Y_{\rm{e}}\, = 0.10 - 0.20$; (2) $Y_{\rm{e}}\, = 0.20 - 0.30$; (3) $Y_{\rm{e}}\, = 0.30 - 0.40$.
The first two of our models correspond to the viewing angle more towards equatorial plane,
whereas the last model corresponds to the abundances expected for the viewing angle more towards the polar direction.
Note that the abundances are derived from \citet{Wanajo14} taking flat mass distribution for each value in the $Y_{\rm{e}}$ range (see \ar{fig:abun}).
The percentage of the lanthanide abundance in the ejecta for the three different models are $\sim 21\%$, $\sim 4.8\%$, and $\sim 0\%$, respectively.
The total ejecta masses for the models are fixed at $M_{\rm ej}\, =\,0.01M_{\odot}$.
Note that model 3 is the same as the fiducial model used in \ct{Smaranikab20},
but with lower ejecta mass ($M_{\rm ej}\, =\,0.01M_{\odot}$ instead of $M_{\rm ej}\, =\,0.05M_{\odot}$).

Our choice of models in 1D are relatively simple.
In reality, the observer from the polar direction would observe the emission from the lanthanide-free disk wind ejecta
surrounded by the faster moving lanthanide-free polar dynamical ejecta.
Similarly, the observer from the equatorial direction would observe the emission from the lanthanide-free disk wind ejecta surrounded by
the lanthanide-rich, faster moving tidal dynamical ejecta.

To study the effects of the density and the abundance structure in the ejecta,
we also include additional models with a continuous thin outer layer with a fixed mass of $M_{\rm{out}}\,=\,0.001M_{\odot}$, moving with velocity $v\,>\,0.2c$.
The layer has a steeper density structure $\rho \propto r^{n^{'}}$.
We take $n^{'} = -10$, since such profiles show good fit with the last kilonova AT2017gfo associated with GW170817 \ctp{kasen17}.
In this case, the maximum outer velocity is $v\,\sim$ 0.33c.
The electron fraction of the inner layer ejecta ($v\le 0.2c$) are fixed to be $Y_{\rm{e}, in}\, = 0.30 - 0.40$,
but the same for the outer layer ejecta ($0.2c< v \le 0.33c$) are varied
and are assumed as different models: (4) $Y_{\rm{e, \,out}}\, = 0.10 - 0.20$; (5) $Y_{\rm{e,\, out}}\, = 0.20 - 0.30$; and (6) $Y_{\rm{e, \,out}}\, = 0.30 - 0.40$.
These models are the more realistic situation for the viewing angles towards the equatorial (models 4 and 5) and the polar direction (model 6).
We summarize all the models in \ar{tab:model_gray}.

To calculate the light curves, we use a time- and wavelength-dependent Monte Carlo radiative transfer code \ctp{Tanaka13, Kawaguchi18}.
The code calculates the multi-color light curves and spectra for a given a density structure and electron fraction
($Y_{\rm{e}}$) distribution assuming the homologously expanding motion of the ejecta.
The radioactive heating rate of the $r$-process nuclei is calculated according to $Y_{\rm{e}}$, by using the results from \citet{Wanajo14}.
The code adopts a time-dependent thermalization factor from \citet{Barnes16}.
Our simulation considers the wavelength range of $\lambda \sim 100 - 35000\, \rm \AA$.

Here we want to mention that the radiative transfer becomes infeasible if we use the complete linelist including the highly ionized lanthanides.
This is because the number of transitions is extremely high.
Hence, to make the linelist for the lanthanide ions, we include reduced number of transitions for the ionization $>$ V.
For this purpose, we follow a method introduced by \ct{Smaranikab22}.
We randomly select the transitions from the original linelist by keeping the statistical properties the same.
Following \ct{Smaranikab22}, we use $0.1\%$ of the total linelist to perform radiative transfer simulations
since such reduced linelist is confirmed to preserve the statistical properties.
More details of the method can be found in \ct{Smaranikab22}.
Note that for the elements with the ionization $\le$ IV, we use the non-reduced complete atomic data from \ct{Tanaka20a}.

Our models assume ejecta masses smaller than that proposed for the kilonova AT2017gfo ($M_{\rm ej}\, =\,0.05M_{\odot}$,
e.g., \cta{Kasliwal17, kasen17, Waxman18, Smaranikab20}).
This is because our work aims to provide conservative estimates for the brightness of the kilonova emission.
Due to this choice, the physical properties and correspondingly the opacity of the ejecta
might be different than the single-element opacity discussed in the \ar{subsec:elem_op}.
Therefore, we first discuss the evolution of the opacity for these models.

\subsection{Opacity evolution} \label{subsec:opev}

\begin{figure*}[t]
\begin{tabular}{c}

\begin{minipage}{0.5\hsize}
\begin{center}
\includegraphics[width=\linewidth]{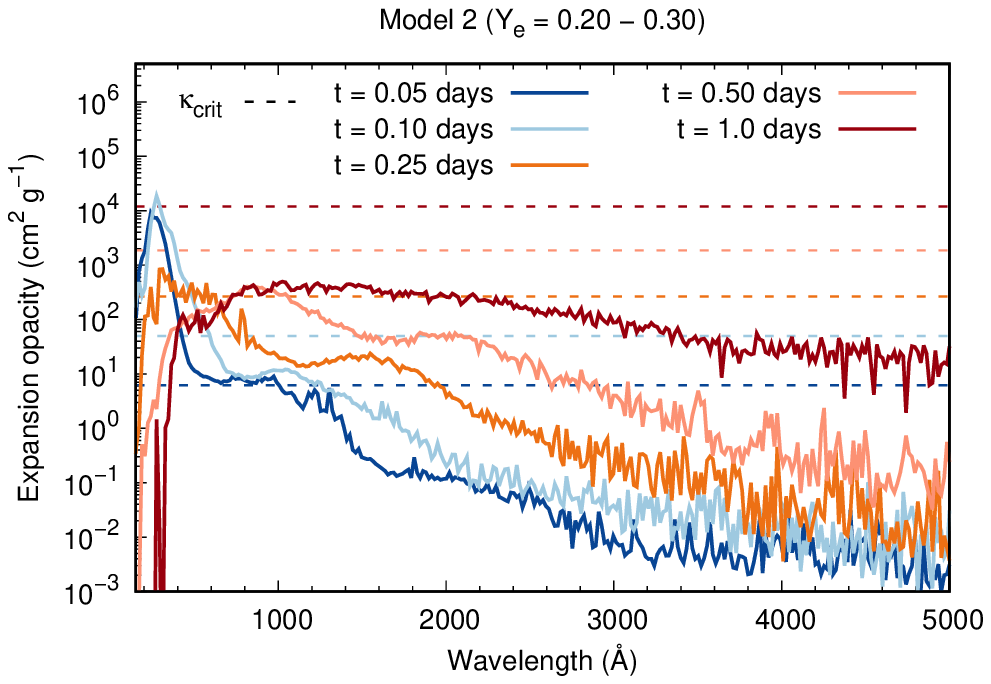}
\end{center}
\end{minipage}

\begin{minipage}{0.5\hsize}
\begin{center}
\includegraphics[width=\linewidth]{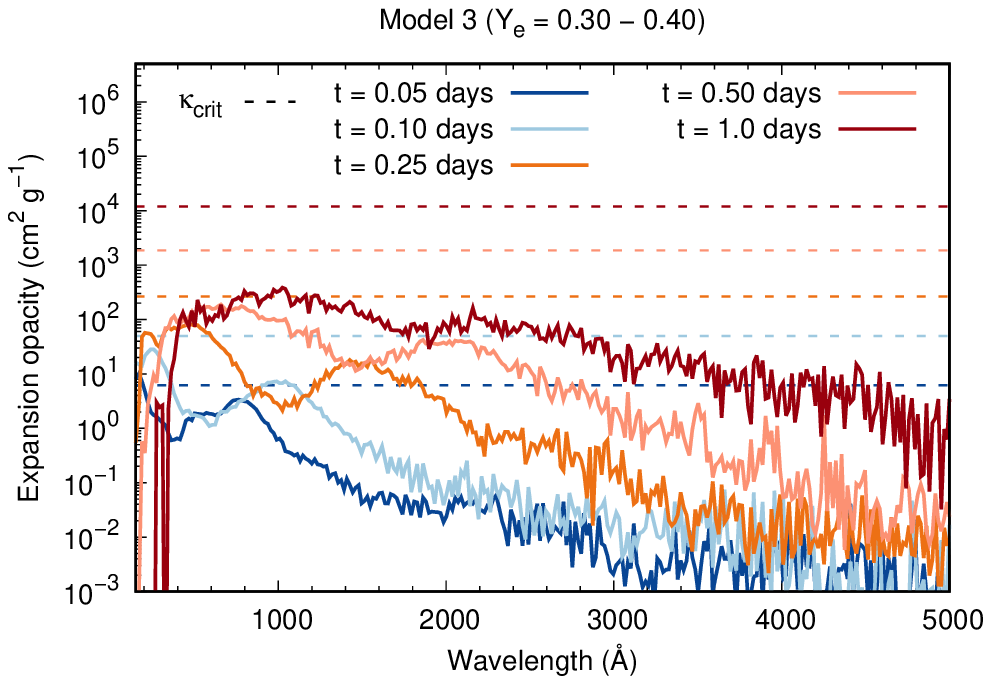}

\end{center}
\end{minipage}

\\

\begin{minipage}{0.5\hsize}
\begin{center}
\includegraphics[width=\linewidth]{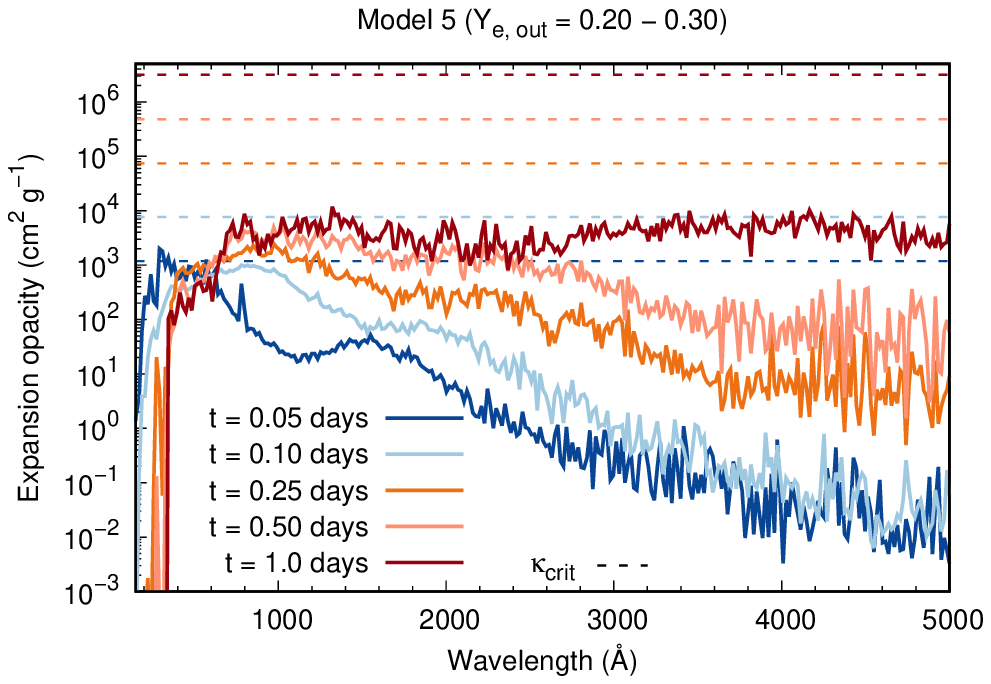}
\end{center}
\end{minipage}

\begin{minipage}{0.5\hsize}
\begin{center}
\includegraphics[width=\linewidth]{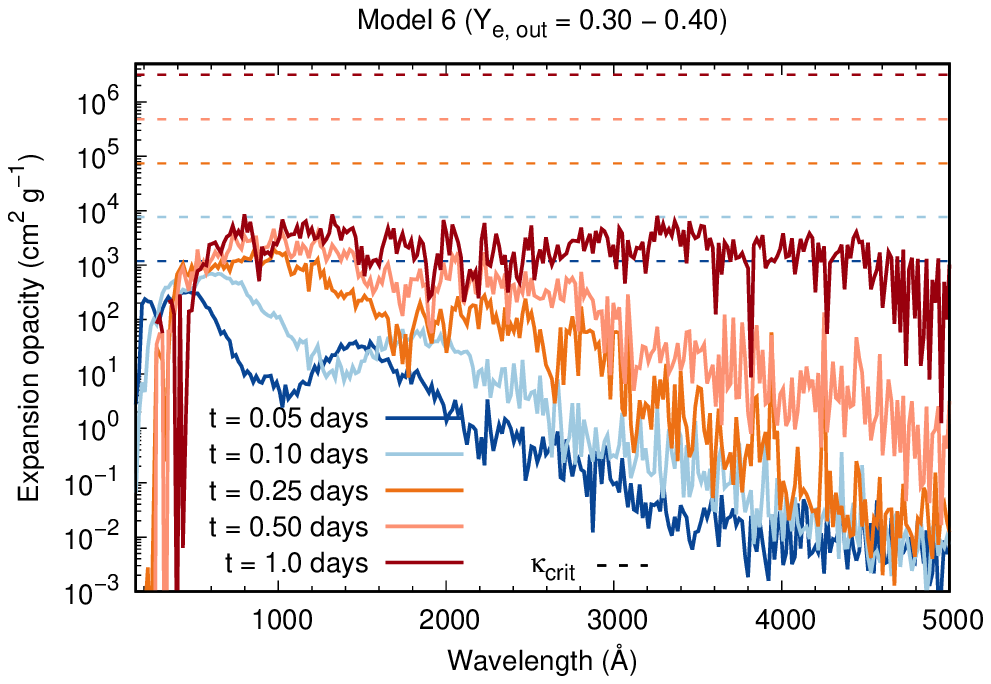}

\end{center}
\end{minipage}

\end{tabular}

\caption{The expansion opacity evolution at the fixed point of ejecta for different models:
$v$ = 0.19c in models 2 and 3 with simple density (top panel); $v$ = 0.33c in models 5 and 6 with continuous thin outer layer density structure (bottom panel).
The ejecta for the models 2 and 5 are lanthanide-rich and that for the models 3 and 6 are lanthanide-free.
The high peak in the opacity is seen only for the lanthanide-rich model with the simple density structure (model 2).
The critical opacities at the fixed ejecta point at different times are also shown (dashed lines).
The expansion opacity approximation is good when the expansion opacities are lower than the critical opacities.
This condition is satisfied for most of the models at different timescales.
The only exception is the opacity for the model 2 at $\lambda \le 1000\, \rm \AA$ at $t < 0.25$ days, when the expansion opacity is underestimated.}

\label{fig:mix_op_ye}
\end{figure*}

\begin{figure*}[t]
\centering
\begin{tabular}{c}
\includegraphics[width=0.5\linewidth]{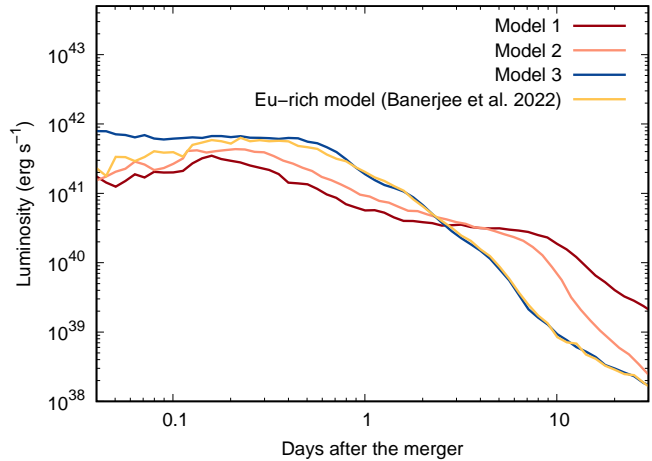}
\end{tabular}
\caption{Bolometric luminosity for different models with simple density and the different abundances (different colored curves).
The ejecta with simple density and single lanthanide elemental abundance ($10\%$ Eu, \cta{Smaranikab22}) is shown for comparison.
If the lanthanides are present in the ejecta, they show unique rising signature at the early time.
Also, for the lanthanide-rich ejecta, the overall light curve is fainter than the lanthanide-free case.}

\label{fig:lc_bol}
\end{figure*}

As the ejecta expand with time, the density and the temperature evolve.
The change in the temperature causes the change in the ionization and correspondingly, changes the opacity. At the early time,
the physical conditions of the ejecta at outermost layer mainly affects the light curve since most of the emission comes from the outermost layer in the ejecta at the early time.

\ar{fig:rhoT_t} shows the physical conditions (density, temperature) of the ejecta at outermost layers in the ejecta for the different models. 
The outermost layer for the models with the simple density (e.g., model 3) and the models with the continuous thin outer layer (e.g., model 6) are different.
Hence, we show the physical conditions at $v$ = 0.19c for the model with the simple density structure (e.g., model 3, orange curves in \ar{fig:rhoT_t})
and at $v$ = 0.33c for the models with the additional thin layers (e.g., model 5, blue curves in \ar{fig:rhoT_t}).
The labels in the right panel of \ar{fig:rhoT_t} shows the peak ionization appearing at different temperature.

The density and the temperature in the outer layer of the ejecta for the simple ejecta model (e.g., model 3)
varies from $10^{-10}\, \rm g \, cm^{-3}$ to $10^{-14}\, \rm g \, cm^{-3}$ and $T\sim 90,000$ K to $T\sim 10,000$ K, respectively,
during the time from $t\sim 0.05$ days to $t\sim 1$ day. The corresponding ionization varies from X to II. 
The outermost layer of the ejecta for the model with a thin layer (e.g., model 6) has a lower density and correspondingly, lower temperature and ionization.
The other models with same density structure show almost the same trends for the temperature (and hence the ionization) evolution in the outermost layers,
although there might be slight variation due to the differences in the compositions in different models.

The expansion opacity evolves with the changes in the physical conditions (temperature and density) in the ejecta.
The top panel of the \ar{fig:mix_op_ye} shows the evolution of the expansion opacity at the outer layers ($v\sim 0.19c$) for models 2 and 3,
i.e., the simple models with and without lanthanides.
Note that we show the opacity only for one lanthanide-rich model (model 2, with lanthanide fraction is $4.8\,\%$)
since the evolution is similar in the other lanthanide-rich model (model 1, lanthanide fraction is $21\,\%$),
except for the fact that the opacity is higher due to the higher lanthanide abundances.

The opacity for the mixture of the elements in the ejecta show strong wavelength dependence similar to the opacity for the individual elements.
The opacity for the ejecta containing lanthanides show extremely high value ($\kappa_{\rm exp} \sim 10^4 \, \rm cm^2\,g^{-1}$)
at around $\lambda \le 1000\, \rm \AA$ at $t\sim 0.05 -0.1$ days, which is much higher than the expansion opacity for the lanthanide-free ejecta
(the peak value of $\kappa_{\rm exp} \sim 1 - 10 \, \rm cm^2\,g^{-1}$ at $\lambda \le 1000\, \rm \AA$) at the same time.
This is because at around $t\le 0.1$ days, the outer ejecta reaches the temperature range ($T \ge 50,000$ K, \ar{fig:rhoT_t})
suitable for the lanthanides to be ionized at $\ge$ VII, which show maximum energy level density (see \ar{sec:atom_res}).

As time progresses, the wavelength-dependent opacity in the outer layer in the ejecta changes due to the changes in the ionization.
For example, the peak in the far-UV wavelengths ($\lambda \le 1000\, \rm \AA$) for the lanthanide-rich ejecta disappears with time as the temperature,
and hence the ionization, decreases.
In addition, the opacity values shift towards higher value because of the changes in the density towards the lower value (see \ar{eqn:kexp}).

The bottom panels of the \ar{fig:mix_op_ye} show the opacity in the outermost layer ($v\sim 0.33c$) for models 5 and 6,
where the lanthanide-rich and lanthanide-free continuous thin outer layer is present.
Interestingly, for the lanthanide-rich ejecta (model 5), the signature high opacity peak at the far-UV wavelengths ($\lambda \le 1000\, \rm \AA$) is not observed. 
This is because the outermost layer is relatively less dense and the temperature is lower than that at $v$ = 0.19c in the models with the simple density structure,
i.e., models 1, 2, 3.
Hence, the temperature range, where the lanthanides are highly ionized ($\ge$ VII), is passed before $t <0.03$ days, the earliest epoch in our simulation.
Similarly, the opacity in the outermost layer for the lanthanide-free models can be understood.

In the recent study to calculate the kilonova signal for the equal-mass binary merger by \ct{Combi23a},
they find the high opacity peaks for lanthanides do not appear in the early time ($t\sim 1$ hour) for the ejecta at the high velocity tail.
Interestingly, our results for the model with lanthanide-rich thin outer layer (model 5),
show the behavior similar to their result despite of a different ejecta distribution.
This is mainly due to the fact that the total mass of the ejecta considered in both the simulations are comparable
(within a factor of few). 
However, for different merger conditions, e.g., a merger of binaries with higher mass ratios, mass of the dynamical ejecta might be higher \ctp{Fujibayashi22},
resulting in higher temperature in the outer layer of the ejecta until $t\sim$ a few hours.
In such a case, the high opacity peaks of lanthanides may significantly affect the opacity and the light curves.







Finally, we discuss the limitations of our opacity calculations. We use expansion opacity approximation for our opacity calculations.
However, if the expansion opacity is beyond the critical opacity at a particular density and temperature, then the opacity is underestimated (\ar{sec:op}).
We show critical opacities at different times as dashed lines in \ar{fig:mix_op_ye}
to assess whether the expansion opacity provides the good approximation at different times.

Our results show that the expansion opacity provides the good approximation over the entire wavelength range considered across different epochs
for all the realistic models with thin continuous outer layer (models 4, 5, 6, \ar{fig:mix_op_ye}).
In addition, the expansion opacity provides good approximation for the model with simple density stucture and with no lanthanide abundances (model 3).
The only exception occurs for the models where the ejecta have the simple density structure and the lanthanide-rich abundances (e.g., model 2)
at the earliest epoch ($t < 0.25$ days) at far-UV wavelengths ($\lambda \le 1000\, \rm \AA$), when the expansion opacity exceeds the critical opacity.
Hence, the opacity is underestimated for models 1 and 2 at $t< 0.25$ days at far-UV ($\lambda \le 1000\, \rm \AA$).
However, even in such case, we can calculate the effective upper limits of the light curves at far-UV,
whereas the light curves at longer wavelengths are expected to be unaffected.

Here we also note that in our calculation, we neglect the actinides ($Z = 89 -103$) 
while calculating the abundance pattern for the ejecta with lower electron fraction. Hence, the overall opacity might be underestimated.
For example, the calculation from \ct{Wanajo14} show that if the electron fraction in the ejecta is $Y_{\rm e} = 0.10 - 0.20$,
the ejecta can contain up to $2\,\%$ of actinides ($Z = 89 -103$). 
Hence, our results of atomic opacity might be underestimated for the models including the ejecta with $Y_{\rm e} = 0.10 - 0.20$, such as models 1 and 4.

\begin{figure*}[t]
\centering
\begin{tabular}{c}
\includegraphics[width=0.5\linewidth]{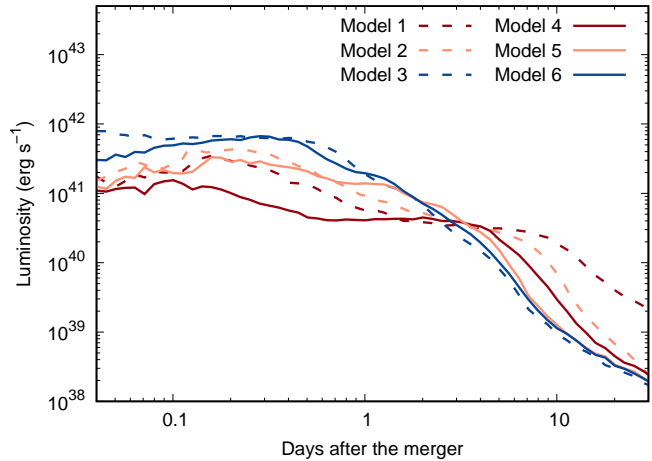}

\end{tabular}
\caption{Bolometric luminosity for different models with simple density (dashed lines) and the models with the additional thin layer (solid lines).
Different colored curves represent the models with different compositions.}
\label{fig:lc_bol_ye}
\end{figure*}


\subsection{Bolometric light curves}\label{sec:lbol}

\begin{figure*}[t]
  \begin{tabular}{c}  
    
    \begin{minipage}{0.5\hsize}
      \begin{center}
        \includegraphics[width=\linewidth]{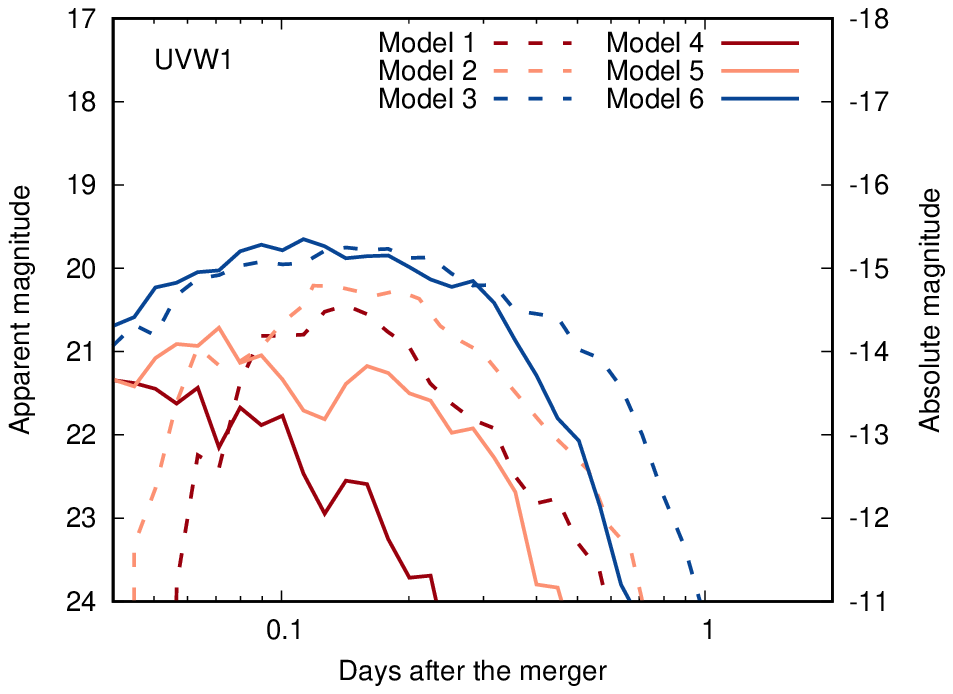}
      \end{center}
    \end{minipage}
   
     \begin{minipage}{0.5\hsize}
      \begin{center}
        
        \includegraphics[width=\linewidth]{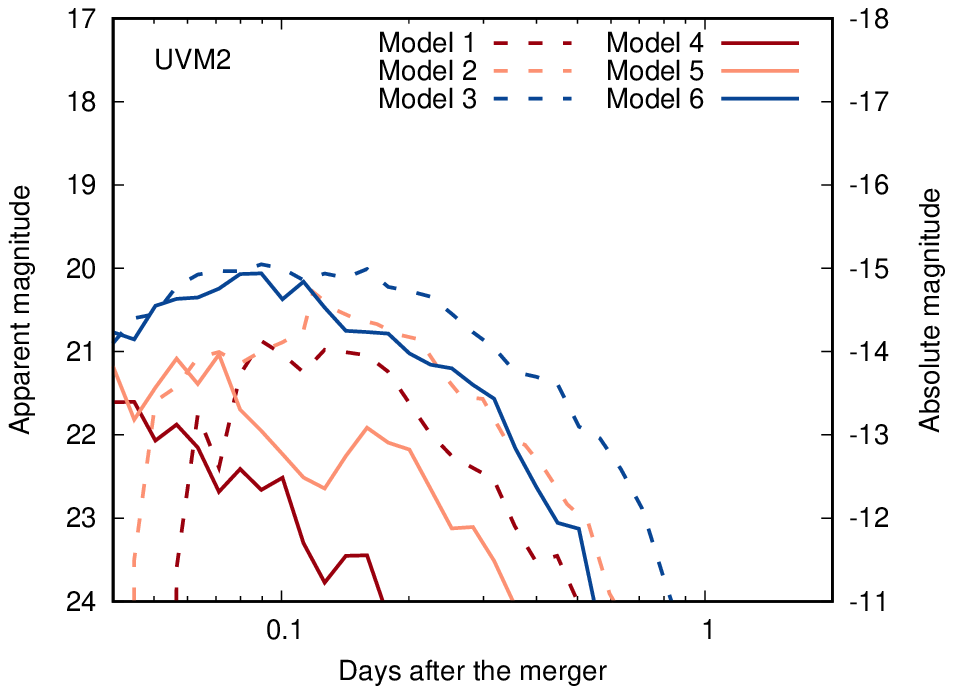}
      \end{center}
    \end{minipage}
  
 \\
\begin{minipage}{0.5\hsize}
      \begin{center}
        \includegraphics[width=\linewidth]{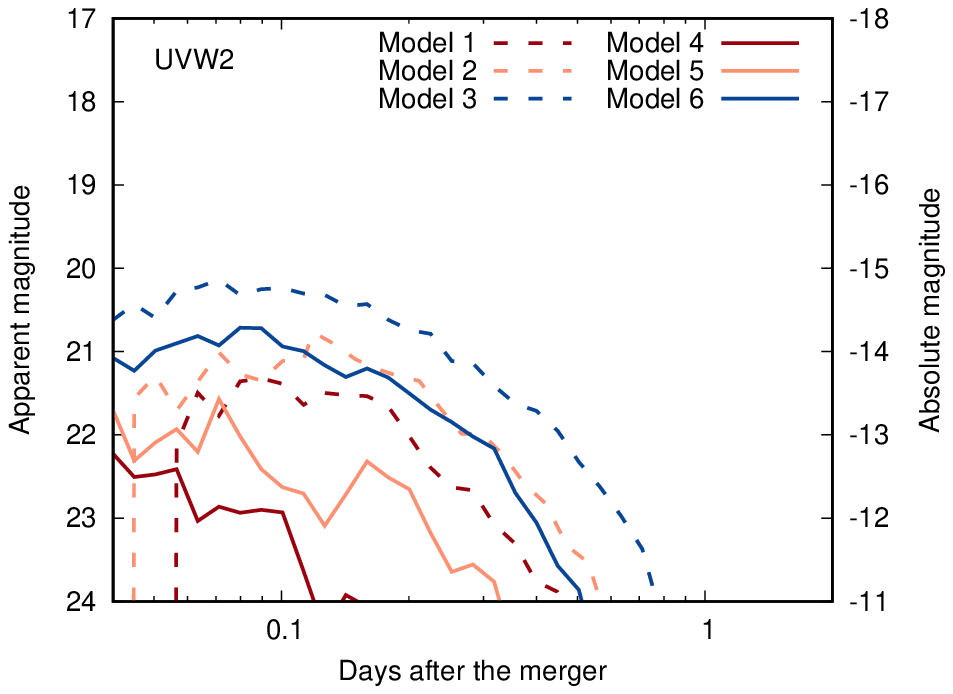}
      \end{center}
    \end{minipage}

\end{tabular}
        \caption{Comparison of UV magnitudes between different models for a source at a distance 100 Mpc.
The magnitudes are shown for three \textit{Swift} filters UVW2, UVM2, and UVW1 with the mean wavelengths 2140 $\rm \AA$, 2273 $\rm \AA$,
and 2688 $\rm \AA$, respectively \ctp{Roming05}.
The light curves are fainter at $t \sim 0.1$ day if the lanthanides are present in the ejecta or if the thin outer layer is present.}

    \label{fig:uv_ye}
\end{figure*}

\begin{figure*}[t]
  \begin{tabular}{c}  
    
    \begin{minipage}{0.5\hsize}
       \begin{center}
        \includegraphics[width=\linewidth]{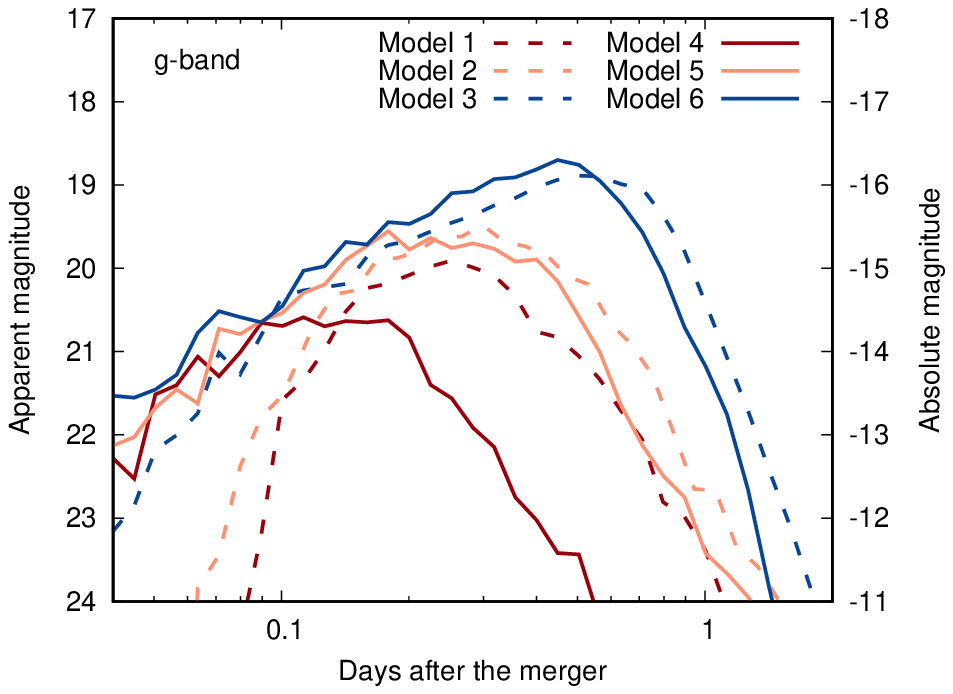}
      \end{center}
    \end{minipage}
   
     \begin{minipage}{0.5\hsize}
      \begin{center}
        
        \includegraphics[width=\linewidth]{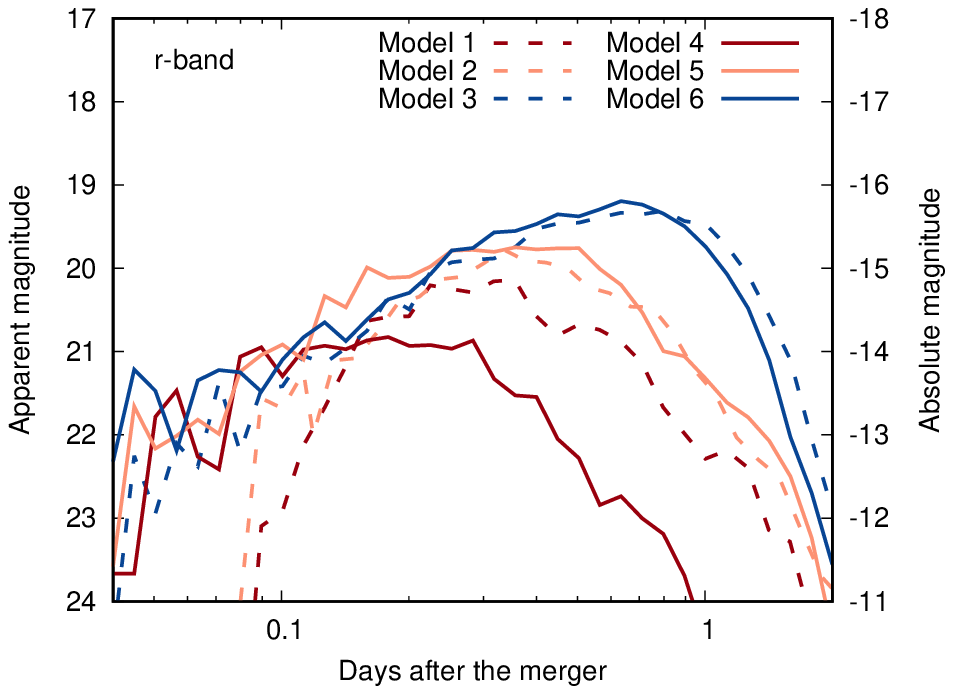}
      \end{center}
    \end{minipage}
  
 \\
\begin{minipage}{0.5\hsize}
      \begin{center}
        \includegraphics[width=\linewidth]{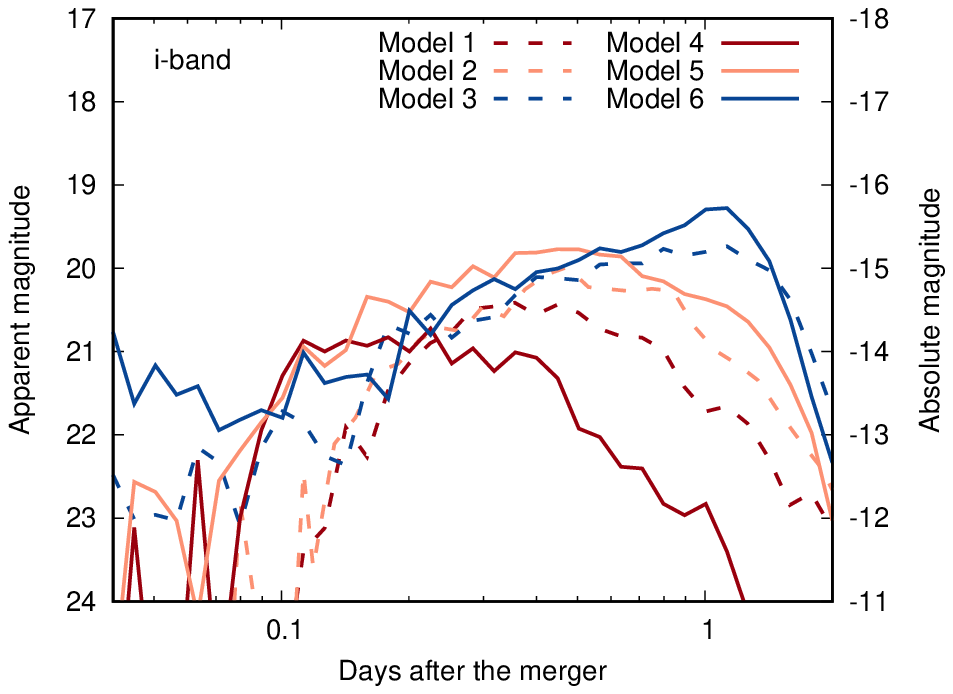}
      \end{center}
    \end{minipage}

\begin{minipage}{0.5\hsize}
      \begin{center}
        \includegraphics[width=\linewidth]{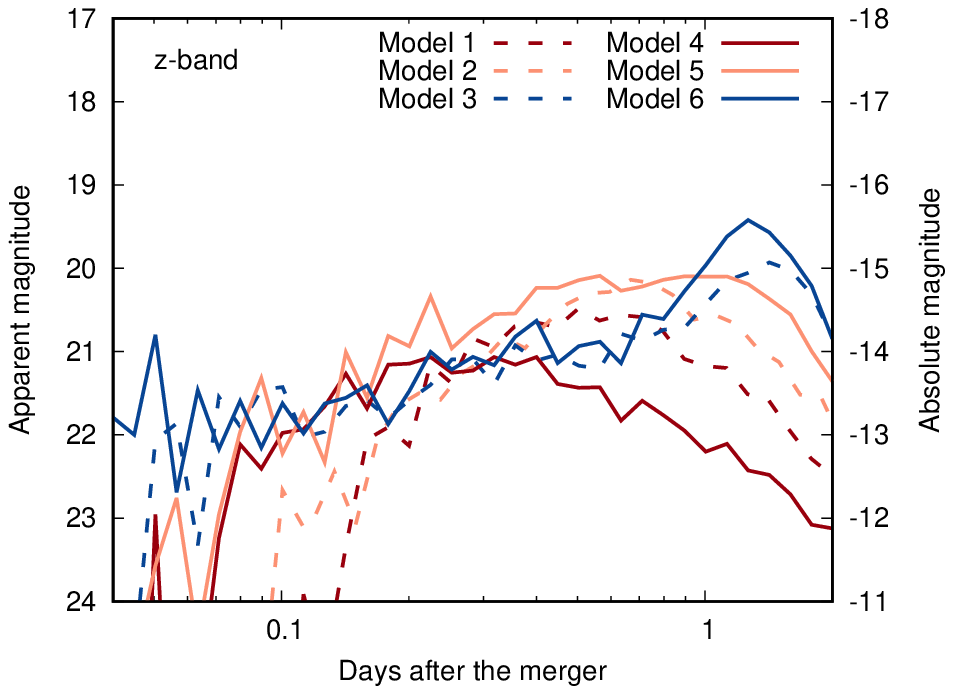}
      \end{center}
    \end{minipage}

\end{tabular}
  \caption{Comparison of magnitudes in four optical filters in $g$-, $r$-, $i$-, $z$-bands for different models for a source at a distance 100 Mpc.}
   
  \label{fig:oir_ye}
\end{figure*}

\ar{fig:lc_bol} shows the bolometric luminosities for different models with the simple density structure (models 1, 2, 3).
Additionally, the luminosity for the ejecta with the same density structure but the composition consisting of the mixture
of a single lanthanide Eu ($Z = 63$, with abundance $\sim 10\%$) and the light $r$-process elements
(corresponding to the electron fraction $Y_{\rm e}$ = 0.30 $-$ 0.40) is shown for comparison.
Note that this is the same model adopted in the \ct{Smaranikab22}, except for the lower ejecta mass considered in this work.

The bolometric luminosities for lanthanide-rich models (models 1 and 2, maroon and orange curves, \ar{fig:lc_bol})
are fainter in comparison with the luminosities of the lanthanide-free model (blue curve in model 3, \ar{fig:lc_bol}).
The bolometric luminosities for the lanthanide-rich ejecta (models 1, 2) reach $\sim L_{\rm bol}\sim 3 - 4 \times10^{42}\,\rm erg\,s^{-1}$ at $t\sim 0.2$ days,
where the higher luminosity is for the lower fraction of lanthanides (model 2) in the ejecta.
On the other hand, the bolometric luminosities for the ejecta with no lanthanides (model 3) reach $L_{\rm bol}\sim 8 \times10^{42}\,\rm erg\,s^{-1}$ at $t\sim 0.2$ days.
The faintness of the light curves in lanthanide-rich models are observed at the later epochs as well.
This is because the presence of the lanthanides makes the opacity in the ejecta higher at all epochs (see \ar{fig:mix_op_ye}),
consequently making the light curves fainter.

The bolometric luminosities for the lanthanide-free ejecta (model 3) are relatively flat at times $t < 1$ day,
whereas that for lanthanide-rich models (models 1, 2) show distinct signature at the same epochs (\ar{fig:lc_bol}).
For instance, the light curves for the lanthanide-rich models (models 1, 2) rise until $t\sim 0.2$ days and decrease afterwards.
Therefore, the trend in the bolometric luminosities before $t\sim 1$ day can act as a distinct signature to infer the presence of the lanthanides in the ejecta.

The shape of the light curves for the lanthanide-rich ejecta at early time can be understood by the opacity evolution in the outer layer of the ejecta.
At $t < 0.1$ days, the temperature of the outermost layer provides the suitable condition ($T \le 70,000$ K)
so that the opacity of the lanthanide-rich ejecta reaches the maximum opacity (see \ar{subsec:opev}).
Such a rise in the opacity in the outermost layer causes the luminosity drop at $t < 0.1$ days.
At around $t\sim 0.2$ days, the opacity in the outer layer of the ejecta decreases with the change in the temperature and ionization,
causing the rise in the luminosity. This work, for the first time, provides the light curves for the realistic abundance pattern of the lanthanide-rich ejecta.

 Note that previously \ct{Smaranikab22} show the effect of the lanthanides in the ejecta is to produce the distinct signature on the light curve.
 However, their model consists of a single lanthanide (e.g., Eu)  instead of the mixture of lanthanides.
 We show that such single lanthanide-rich model (Eu-rich model, similar to \cta{Smaranikab22}, but with lower mass)
 can reproduce the overall trend of the early luminosity at $t < 0.2$ days as the model with the mixture of lanthanides (\ar{fig:lc_bol}).
 However, the absence of second half of the lanthanides overestimates the luminosity beyond $t > 0.2$ days.

The presence of a continuous, thin outer layer in the ejecta (models 4, 5, 6)
makes the early luminosity fainter than that of the cases with the simple density structure (e.g., models 1, 2, 3, \ar{fig:lc_bol_ye}).
For example, the light curves for the models with the lanthanide-free inner and outer ejecta (model 6)
are fainter than those of the models with simple density structure containing no outer layer ejecta (model 3, \ar{fig:lc_bol_ye}).
This is in agreement with the previous works by \ct{kasen17, Smaranikab20}.
In the case where the outer layer contains lanthanides (e.g., model 5), the light curves are even fainter (\ar{fig:lc_bol_ye}).
Note that the presence of the lanthanides in the fast-moving thin outer layer in the ejecta
produces light curves fainter than the lanthanide-rich model with the simple density structure (e.g., model 2).
Hence, there are possibilities of deducing the density structure as well as compositions from the shape of the light curves at the early times.

The lower luminosity for the models with thin continuous outer layer (e.g., model 5) is mainly due to the lower density,
and correspondingly, lower radioactive power deposited in the outer layer.
The photons escape only from the outer layer at the early time, and hence, lower deposited power at the outer layer makes the light curves fainter.
Note that the lower density in the outer layer is also the main reason for lower luminosity even for the models with the lanthanide-rich thin outer layer (e.g., model 5),
and not the high opacity of the ejecta in the presence of lanthanides.
In fact, in this case, the opacity in the outer layer is not extremely high even at the earliest epochs due to the
lower density and lower temperature in the outer layer in such models (\ar{fig:rhoT_t}, \ar{fig:mix_op_ye}).

Here note that the temperature of the ejecta is mass dependent. If the ejecta mass (and correspondingly the outer layer mass) is high,
the ejecta temperature is higher than that of the ejecta with lower mass.
With the change in the mass and the temperature in the outer layer, the shape of the overall luminosity is expected to be affected.


\begin{figure*}[t]
\begin{tabular}{c}  

\begin{minipage}{0.5\hsize}
\begin{center}
\includegraphics[width=\linewidth]{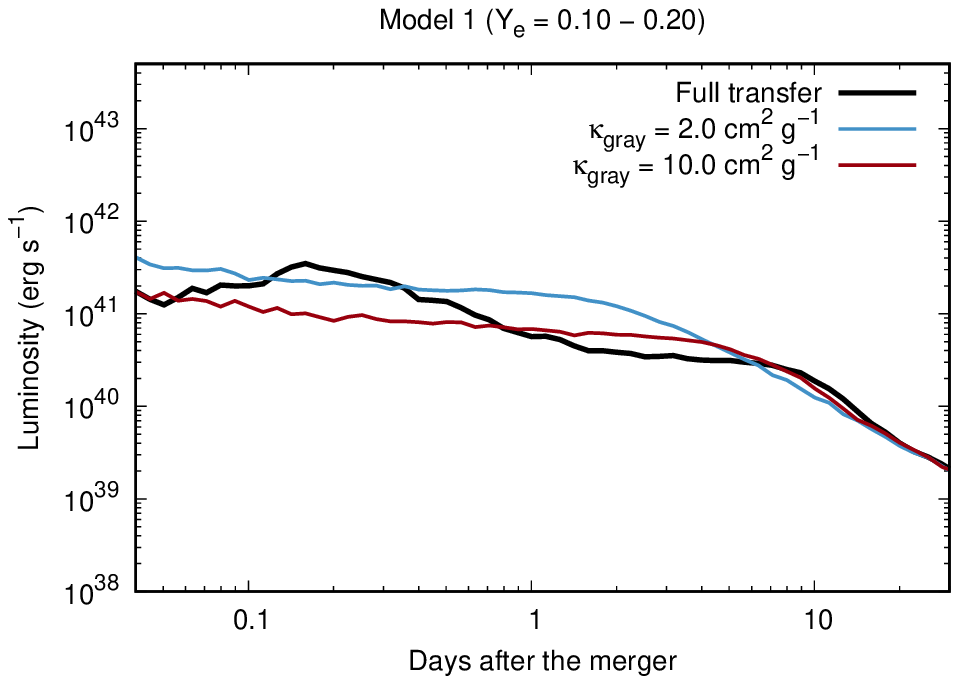}
\end{center}
\end{minipage}

\begin{minipage}{0.5\hsize}
\begin{center}

\includegraphics[width=\linewidth]{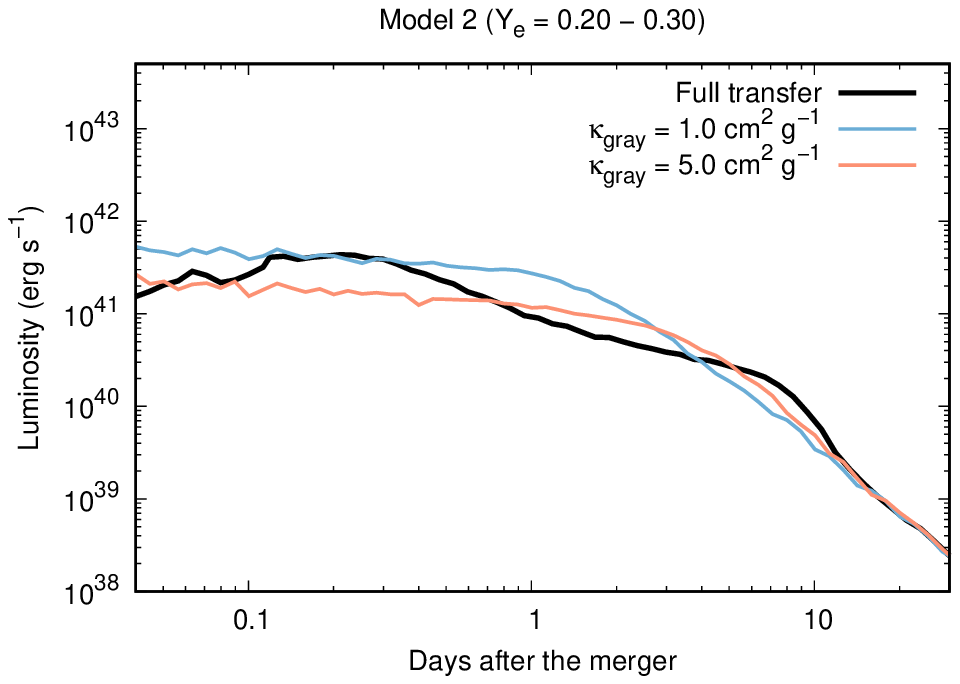}
\end{center}
\end{minipage}

\\
\begin{minipage}{0.5\hsize}
\begin{center}
\includegraphics[width=\linewidth]{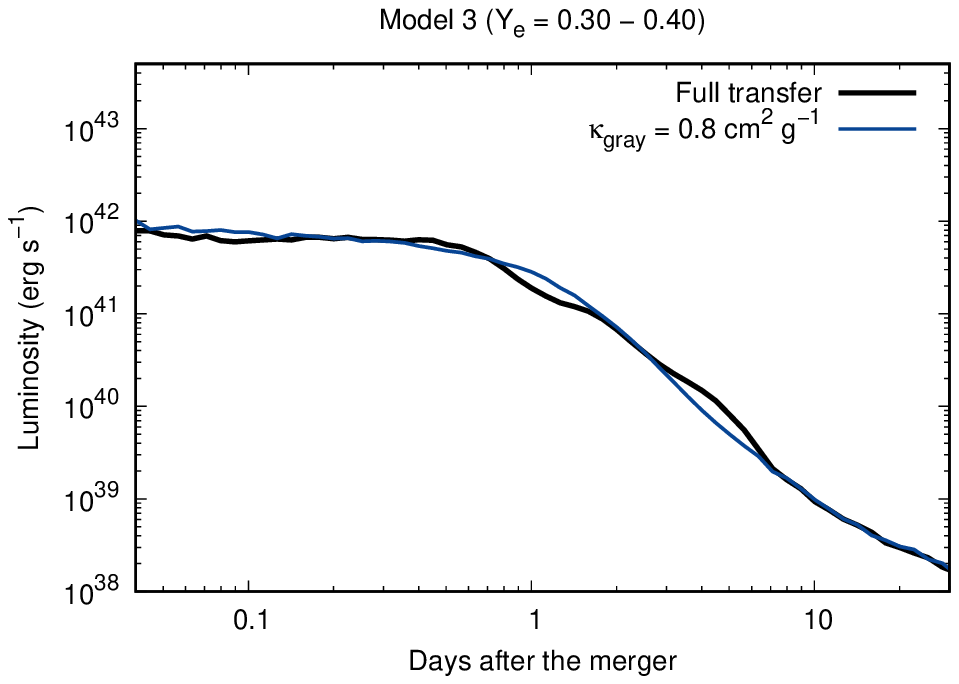}
\end{center}
\end{minipage}

\end{tabular}
\caption{Bolometric light curves for the models with the single density slope.
The different panels represent different abundances in the thin outer layer.
The light curves are compared with the light curves for the same model but with gray opacity.}

\label{fig:gray}
\end{figure*}

\begin{figure*}[t]
\begin{tabular}{c}

\begin{minipage}{0.5\hsize}
\begin{center}
\includegraphics[width=\linewidth]{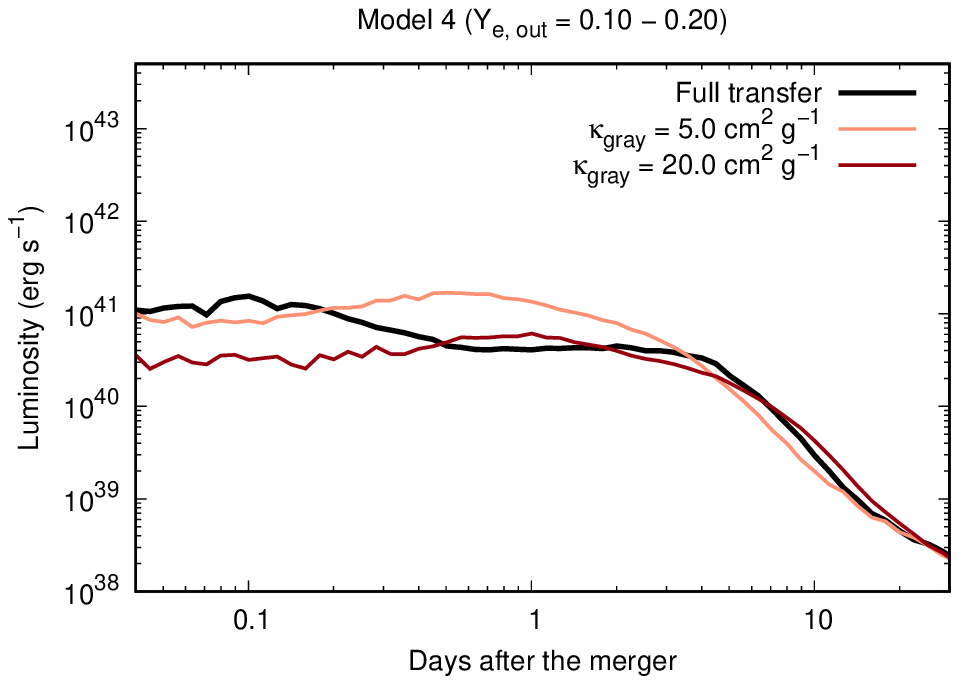}
\end{center}
\end{minipage}

\begin{minipage}{0.5\hsize}
\begin{center}

\includegraphics[width=\linewidth]{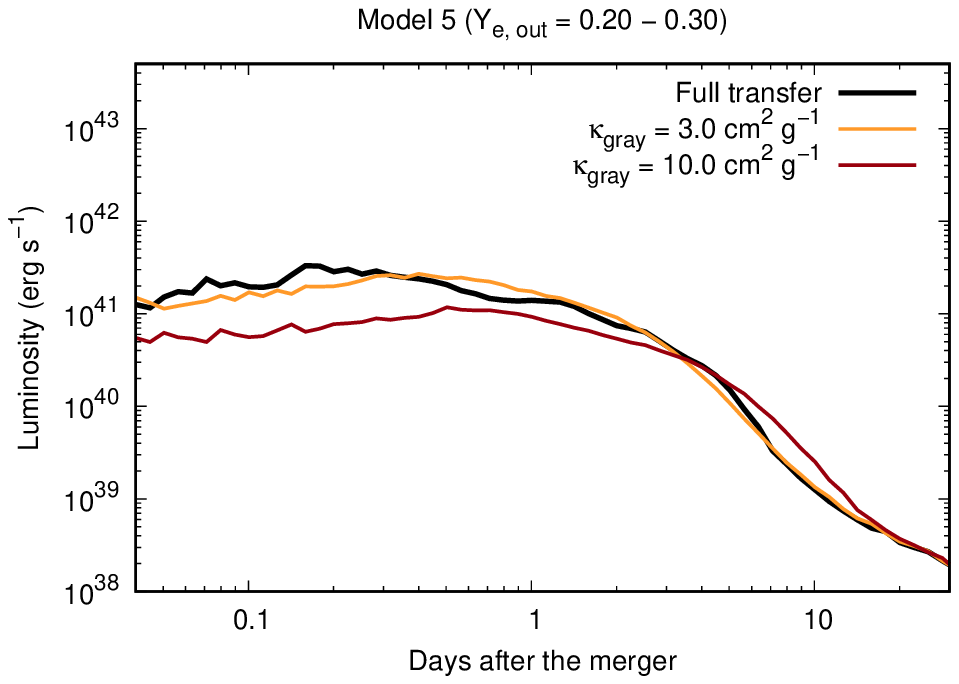}
\end{center}
\end{minipage}

\\
\begin{minipage}{0.5\hsize}
\begin{center}
\includegraphics[width=\linewidth]{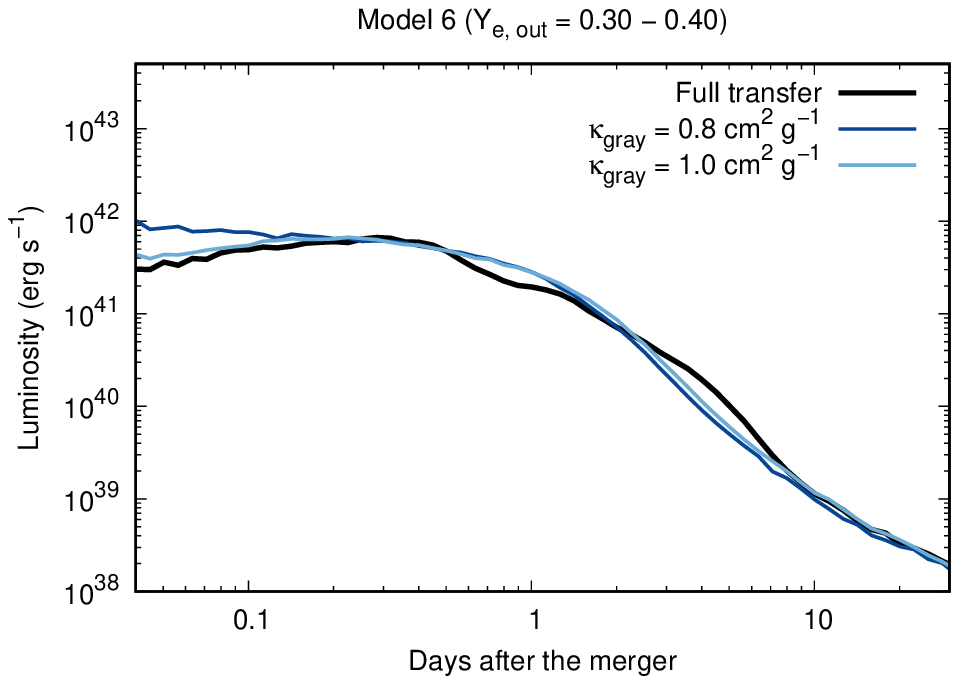}
\end{center}
\end{minipage}

\end{tabular}
\caption{Bolometric light curves for the models with the thin outer layer with steeply declining density slope.
The different panels represent different abundances in the thin outer layer.
The light curves are compared with that for the same model but with the UVOIR transfer with gray opacity.}
\label{fig:gray_ye}
\end{figure*}


\begin{figure*}[t]
\begin{tabular}{c}

\begin{minipage}{0.5\hsize}
\begin{center}
\includegraphics[width=\linewidth]{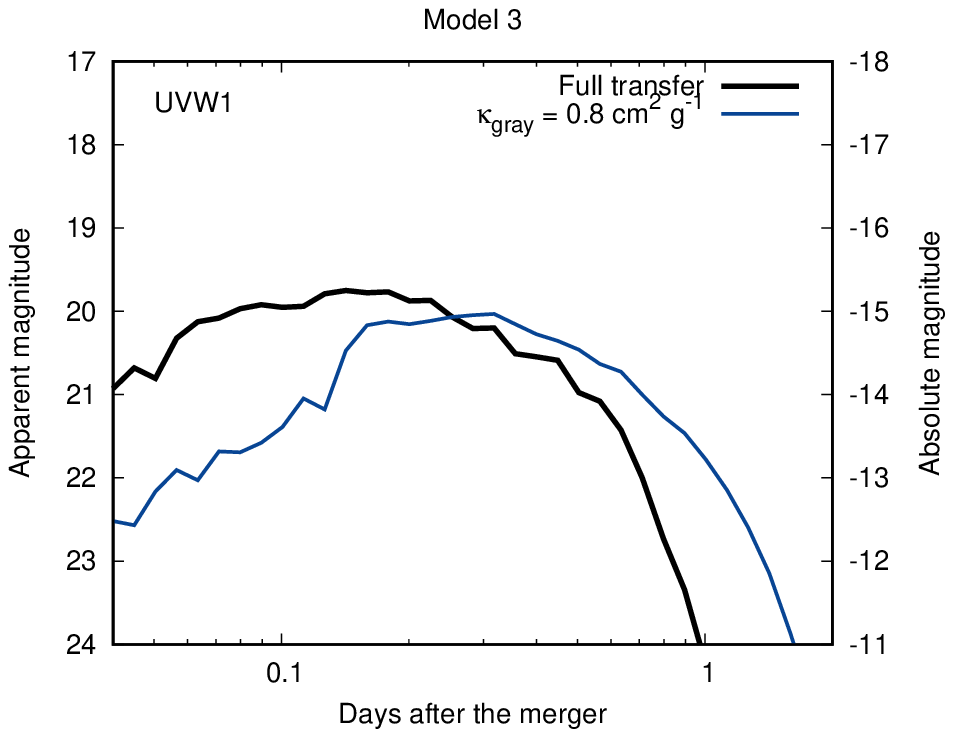}
\end{center}
\end{minipage}
\\

\begin{minipage}{0.5\hsize}
\begin{center}

\includegraphics[width=\linewidth]{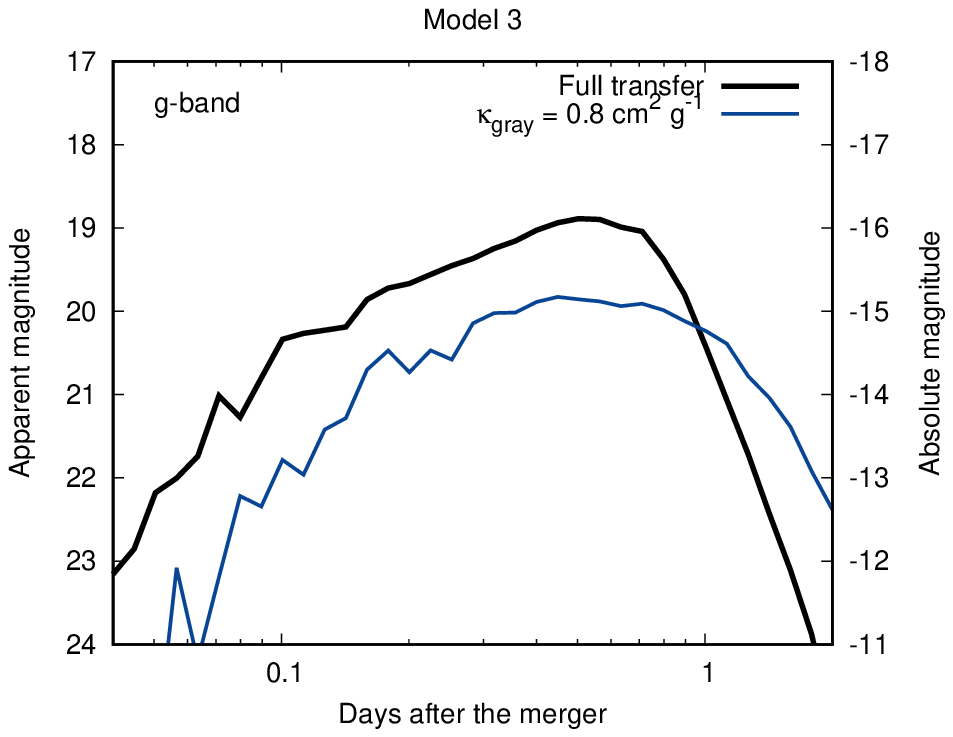}
\end{center}
\end{minipage}

\begin{minipage}{0.5\hsize}
\begin{center}
\includegraphics[width=\linewidth]{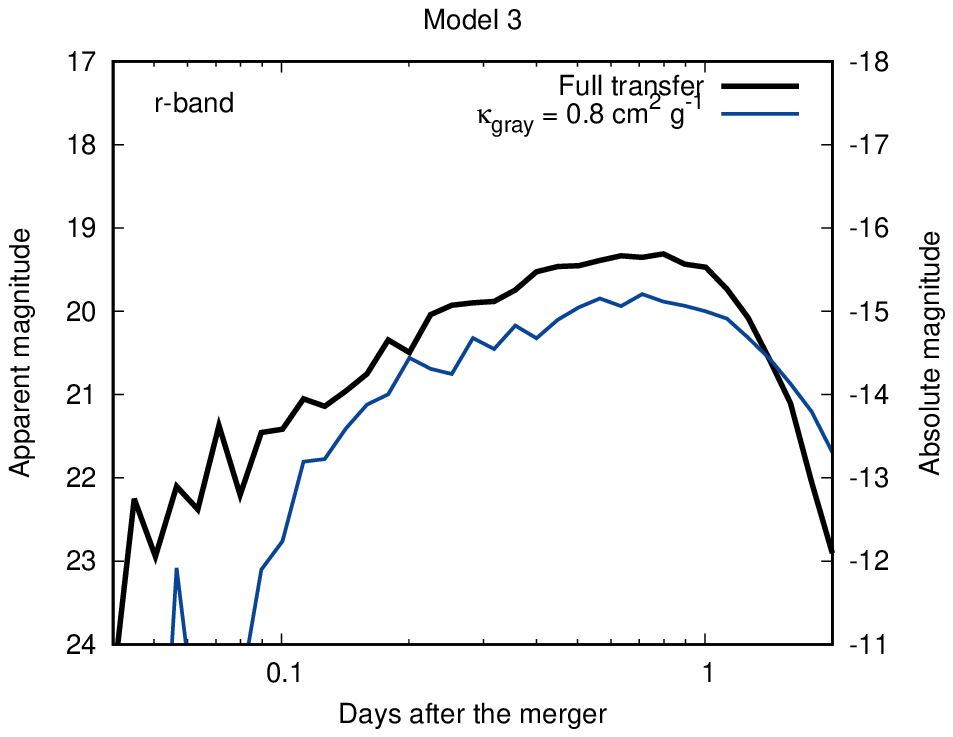}
\end{center}
\end{minipage}

\\

\begin{minipage}{0.5\hsize}
\begin{center}
\includegraphics[width=\linewidth]{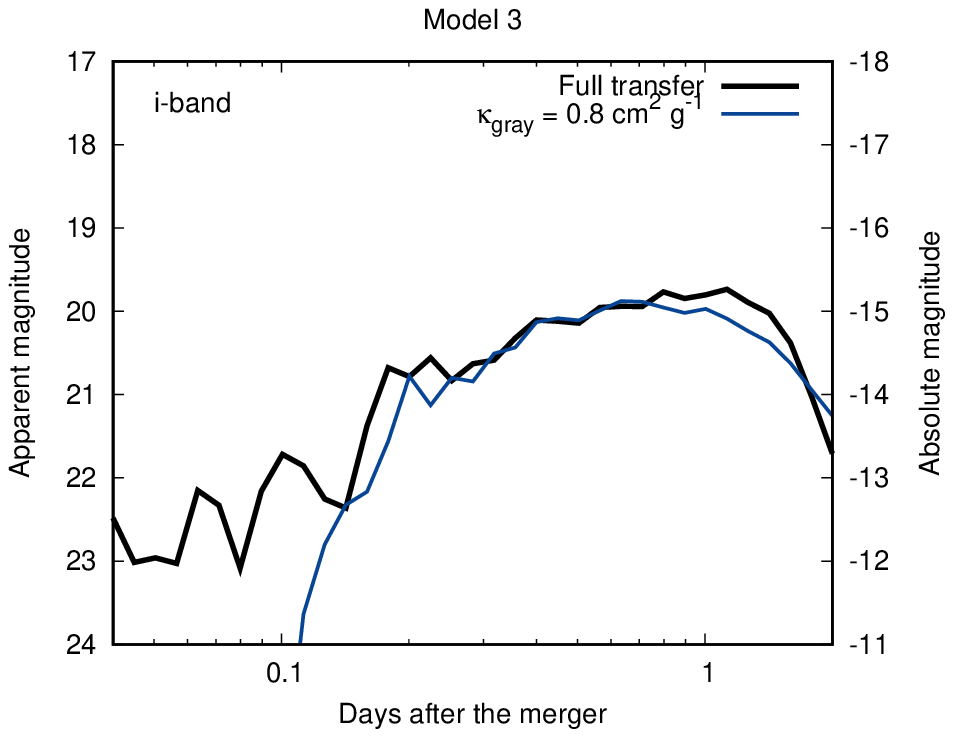}
\end{center}
\end{minipage} 

\begin{minipage}{0.5\hsize}
\begin{center}
\includegraphics[width=\linewidth]{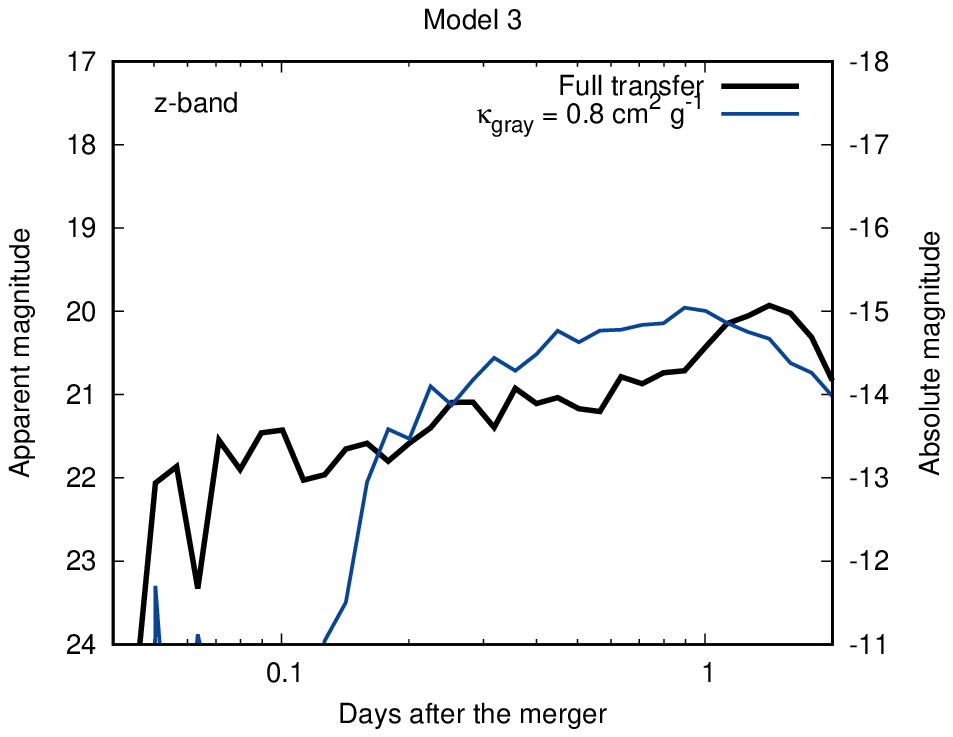}
\end{center}
\end{minipage}

\end{tabular}
\caption{Comparison between the magnitudes in \textit{Swift} filter UVW1 \ctp{Roming05} and the four optical filters in $g$-, $r$-, $i$-, $z$-bands
for the calculations assuming the UVOIR transfer with multi-wavelength opacities (black) and the gray opacities (blue) for model 3,
i.e., the model with the simple density and the lanthanide-free abundances. The source is asssumed to be at a distance 100 Mpc.
The multi-color light curves cannot be reproduced with those from the UVOIR transfer calculations with the gray opacity.}
\label{fig:gray_ye_multi}
\end{figure*}

\subsection{Multi-color light curves}\label{sec:lmulti}

\ar{fig:uv_ye} and \ar{fig:oir_ye} show the UV magnitudes in the three different \textit{Swift} UVOT filters
(UVW2, UVM2, and UVW1 with the mean wavelengths 2140 $\rm \AA$, 2273 $\rm \AA$, and 2688 $\rm \AA$, respectively, \cta{Roming05})
and the four optical filters ($g$-, $r$-, $i$-, $z$-bands) for a source at 100 Mpc.
The models with the simple density structure and that with the thin, continuous outer layer are shown with the dashed and solid lines, respectively.

For the models with the simple density (models 1, 2, 3), our results show that the UV brightness varies from $\sim 22 - 20$ mag at $t\,\sim 0.1$ days,
fainter in the lanthanide-rich models (models 1, 2). For the $g$-, $r$-, $i$-, $z$-filters, the early luminosity varies from  $\sim 21 - 19$ mag at $t\,\sim 0.1$ days.
The models with thin outer layer (models 4, 5, 6) are at the fainter magnitudes, similar to the behavior of the bolometric light curves.

The UV light curves are the promising observable signal at early time.
If the kilonova is discovered early enough so that the prompt observation can be started, then the signals
can be detected with the existing satellite \textit{Swift} (with a limiting magnitude of $\sim$ 22 mag for an exposure time of 1000 s, \cta{Roming05}).
The UV horizon is going to be broadened as many wide-field satellites such as ULTRASAT (limiting magnitude of 22.4 mag for 900 s of integration time, \cta{Sagiv14}),
Dorado (limiting magnitude of 20.5 mag for 600 s of integration time, \cta{Dorsman22}), UVEX (limiting magnitude of 25 mag for 900 s of integration time,
\cta{Kulkarni21}) are upcoming in the next decade. Hence, in the near future, UV counterpart detection probability will increase manifold.

For the optical $g$-, $r$-, $i$-, $z$-filters, the brightness of the kilonovae for the lanthanide-rich ejecta is close to the detection limit
of the existing observing facilities like Zwicky Transient Facility
(ZTF, the limiting magnitudes for ZTF in $g$-, $r$-, $i$-bands are 21.1 mag, 20.9 mag, 20.2 mag, respectively, for the 30 s exposure time, \cta{Dekany20}).
The detection of early kilonova in optical bands seems more feasible for the observing facilities with deeper observing limits,
such as, Dark Energy Camera (DECam, the limiting magnitudes in $i$-, $z$-bands are 22.5 mag, 21.8 mag, respectively, for the 90 s exposure time, \cta{Chase22}),
Subaru-HSC, which can reach a depth of $\sim 24$ mag for $2\times30$ s exposure time (e.g., \cta{Ohgami21, Ohgami23}).
More promisingly, such signals in the optical bands can easily be detected by the upcoming wide-field survey such as Vera Rubin Observatory \ctp{Chase22}.

Here we note that our opacity in the far-UV wavelengths does not provide the correct estimate,
hence our light curves are rather uncertain in far-UV ($\lambda \le 1000\, \rm \AA$) for the lanthanide-rich models with simple ejecta structure.
However, our light curves in the longer wavelengths are likely to be unaffeced by this choice.
Since the detection ranges of the existing instruments discussed here are all beyond 2000 $\rm \AA$ (e.g., Swift, \cta{Roming05}),
our models provide useful predictions for the early kilonova.

\section{Gray opacities to reproduce early kilonova}\label{sec:disc}

Several of the previous studies adopt gray, constant opacities to reproduce the early light curves of the kilonova AT2017gfo associated with GW170817.
This is because of the absence of the detailed opacities in the early time.
Since we perform the multi-frequency transfer to calculate the early light curves for kilonova using the detailed, wavelength-dependent opacity
for almost all the feasible abundances, we now assess how good the gray approximations are at reproducing the early light curves. 
For this purpose, we perform the radiative transfer simulations by adopting the constant opacities of $\kappa_{\rm gray} \sim 0.1 - 20\, \rm cm^2\,g^{-1}$
and compare the results with those with wavelength dependent opacities.

The bolometric luminosity at $t < 1$ day for the model with the simple density structure and lanthanide-free abundances (model 3)
can be reproduced by the gray transfer with the opacity $\kappa_{\rm gray} \sim 0.8\,\rm cm^2\,g^{-1}$ (\ar{fig:gray}).
This is because the outer layer opacity changes gradually at time $t < 1$ day, and hence,
the behavior of the UVOIR transfer with the wavelength dependent opacity can be simulated UVOIR transfer with the gray opacity. 
However, for the similar models but with lanthanide-rich abundances (models 1, 2), the luminosity cannot be well reproduced with a single gray opacity
owing to the rapid changes in the opacity in the outer layer at $t < 1$ day.
Similarly, for the models where the thin outer layer is present, the early luminosities cannot be reproduced (\ar{fig:gray_ye}).

The gray opacity might be a fair approximation to derive the early luminosities if a range of the opacity values are considered instead of a single value.
For example, take model 2, i.e., the simple model with the lanthanide-rich abundance for $Y_e = 0.20 - 0.30$.
The UVOIR transfer with the gray opacity of $\kappa_{\rm gray} \sim 5\,\rm cm^2\,g^{-1}$ reproduces the faintest portion of the light curves at $t \sim 0.05$ days,
whereas the brightest portion of the light curve around $t \sim 0.2$ days can be reproduced by the gray opacity of $\kappa_{\rm gray} \sim 1\,\rm cm^2\,g^{-1}$.
Similarly, gray opacity equivalent for other models can be estimated. The results are summarized in \ar{tab:model_gray}.
However, we want to stress the fact that the detailed trends of the light curves cannot be reproduced with a single gray opacity.

The multi-band light curves cannot be reproduced by the UVOIR transfer with the gray opacity even if the bolometric luminosities show good match.
Take for example, the bolometric and the multi-color light curves for the model 3 with the simple density and the lanthanide-free abundances.
Although the bolometric luminosity shows good match with the light curves with the gray opacity of $\kappa_{\rm gray} \sim 0.8\,\rm cm^2\,g^{-1}$ (\ar{fig:gray}),
the multi-color light curves does not match (\ar{fig:gray_ye_multi}).
Hence, using the wavelength dependent opacity is necessary to explain the multi-band light curves from the observations.

Note that even though the opacity for the lanthanide-rich ejecta shows an extremely high value at the far-UV wavelengths,
the gray opacities required to reproduce the light curves at the early time is not very high.
This is because of the rapid evolution of the spectral peak towards the longer wavelengths due to temperature evolution in the outer layer ejecta.
Hence, the gray opacities to reproduce the light curves resembles the wavelength-dependent opacity at the longer wavelengths, which are not so high.


\section{Conclusions}\label{sec:conc}

To investigate the early kilonova emission from the neutron star merger ejecta,
we perform the atomic opacity calculation for all the elements from La - Ra ($Z = 57 - 88$) ionized to the states V - XI,
which are the typical condition at $t= 0.1$ days (with $T\,\sim \,10^{5}$ K).
This work, together with the previous works, \ct{Smaranikab20, Smaranikab22},
provides the atomic opacities suitable for early time for all the elements from Ca to Ra ($Z = 20 -88$) ionized from V - XI.

Our results show that the opacity varies widely depending on the existing open shell in the ions (\ar{fig:expwav_fdps}).
For instance, the opacities for the lanthanides are exceptionally high,
reaching $\kappa_{\rm exp}\,\sim 10^4 \,\rm{cm^{2}\,g^{-1}}$ at far-UV ($\lambda \le 1000\, \rm \AA$) at the ionizations $\ge$ VII (\ar{fig:elev_vii_xi}).
Similarly, the opacities for the elements with the open $d$-, $p$-, and $s$-shells reach 
$\kappa_{\rm exp}\,\sim 1\,\rm{cm^{2}\,g^{-1}}$, $0.1\,\rm{cm^{2}\,g^{-1}}$, and $0.01\,\rm{cm^{2}\,g^{-1}}$, respectively,
at the same wavelength range (\ar{fig:expwav_fdps}).

Using the new opacity dataset, we perform the radiative transfer simulations to derive the early kilonova light curves using 1D spherical ejecta model
with a power law density structure. In neutron star mergers, the heaviest elements including lanthanides are expected to be distributed near to the equatorial plane,
whereas lighter $r$-process elements are distributed towards the pole or more isotropically.
Because of the different elemental abundances, the opacity is different in different components, introducing the viewing angle dependence in the kilonova light curves.
To understand the light curves from polar and equatorial directions, we assume lanthanide-free and lanthanide-rich abundance of the ejecta, respectively.
Furthermore, we study the effect of a faster moving thin outer layer with different compositions.

We find that in presence of lanthanides in the ejecta, the bolometric luminosities
($L_{\rm bol}\sim 3-4 \times10^{41}\,\rm erg\,s^{-1}$, higher for lower lanthanide fraction in the ejecta, e.g., model 2)
are fainter in comparison with the bolometric luminosities
for lanthanide-free ejecta ($L_{\rm bol} \sim 8\times10^{41}\,\rm erg\,s^{-1}$, \ar{fig:lc_bol}).
For the lanthanide-rich ejecta, there are distinct signature in the early light curves determined
by the evolution of the opacity in the outermost layer in the ejecta (\ar{fig:lc_bol_ye}).
Furthermore, the presence of a thin outer layer suppresses the early luminosity, agreeing with the results of \ct{kasen17, Smaranikab20}.

The UV brightness for a source at 100 Mpc at early time varies from $\sim 22 - 20$ mag,
fainter for the lanthanide-rich ejecta, at $t\,\sim 0.1$ days in \textit{Swift} UVOT filters \ctp{Roming05}.
If the kilonova is discovered early enough so that the prompt observation can be started,
then the UV signals can be detected with the the existing satellite \textit{Swift}
(with a limiting magnitude of $\sim$ 22 mag for an exposure time of 1000 s, \cta{Roming05}),
or the upcoming UV satellites like ULTRASAT (limiting magnitude of 22.4 mag for 900 s of integration time, \cta{Sagiv14}),
Dorado (limiting magnitude of 20.5 mag for 600 s of integration time, \cta{Dorsman22}),
UVEX (limiting magnitude of 25 mag for 900 s of integration time, \cta{Kulkarni21}).

The luminosities in the four optical filters ($g$-, $r$-, $i$-, $z$-bands) appears to vary from
$\sim 21 - 19$ mag at $t\,\sim 0.1$ days for a source at $100\, \rm Mpc$.
The light curves are fainter for the lanthanide-rich models, as in the bolometric luminosities.
Moreover, the presence of the thin outer layers make the light curves even fainter.
For the optical $g$-, $r$-, $i$-, $z$-filters, the kilonovae for the lanthanide-rich ejecta are comparable to the detection limit
of the existing observing facilities like ZTF \ctp{Dekany20}. However, such kilonova might be detecable by the facilities
with deeper observation limit such as DECam \ctp{Chase22}, Subaru-HSC (e.g., \cta{Ohgami21, Ohgami23}). 
More interestingly, such kilonovae are the good targets for the the upcoming wide-field surveys such as Vera Rubin Observatory \ctp{Chase22}.

Finally we mention the limitations in our simulations. Our opacity in the far-UV wavelengths ($\lambda \le 1000\, \rm \AA$) may be underestimated,
and hence, our light curves are possibly affected in far-UV for lanthanide-rich models.
However, the light curves in the longer wavelengths are likely to be unaffeced by this choice since
the detection ranges of the existing instruments are all beyond 2000 $\rm \AA$ (e.g., Swift, \cta{Roming05}; ZTF, \cta{Dekany20}).
Furthermore, our 1D models are relatively simple. The effect of the more realistic multi-dimensional ejecta structure is explored in an upcoming work.

\acknowledgments

Numerical simulations presented in this paper were carried out with
Cray XC-50 at the Center for Computational Astrophysics, National Astronomical Observatory of Japan;
and at the computer facility in the Yukawa Institute for Theoretical Physics (YITP), Kyoto University, Japan.
SB wants to thank K. Kawaguchi, S. Fujibayashi, M. Shibata, J. Barnes, R. Fernandez for discussions in EMMI + IRENA workshop 2022 in GSI, Germany.
SB also wants to thank S. Saito, H. Hamidani, N. Domoto for discussion in Tohoku University.
This reseach was supported by the Grant-in-Aid for Scientific Research from JSPS (19H00694,20H00158,21H04997), JST FOREST Program 291 (Grant Number JPMJFR212Y),
Grant-in-Aid for Scientific Research; MEXT (17H06363), the Grant-in-Aid for JSPS Fellows (22J14199); and the NIFS Collaboration Research Program (NIFS22KIIF005). 
\vspace{5mm}


\bibliography{./references.bib}{}
\bibliographystyle{aasjournal}
\renewcommand{\thetable}{\Alph{section}\arabic{table}}
\renewcommand\thefigure{\thesection\arabic{figure}}

\appendix
\setcounter{table}{0}
\section{The configurations used for calculation}

\begin{center}

\begin{longtable}{llccc}

\caption[]{The configurations and the results from atomic structure calculations.
$N_{\rm level}$, $N_{\rm line}$, and $N_{\rm line}^{*}$ represents the number of energy levels, transitions involving all the energy levels,
and the transitions involving the energy levels below the ionization threshold, respectively.}\\

\hline
\textbf{Ion} & \textbf{Configurations} & \textbf{$N_{\rm level}$} & \textbf{$N_{\rm line}$} &\textbf{$N_{\rm line}^{*}$}\\
\hline 
\endfirsthead
\multicolumn{3}{l}{\tablename\ \thetable\ -- \textit{Continued from previous page}} \\
\hline

\textbf{Ion} & \textbf{Configurations} & \textbf{$N_{\rm level}$} & \textbf{$N_{\rm line}$} &\textbf{$N_{\rm line}^{*}$}\\
\hline
\endhead
\hline  
\multicolumn{3}{r}{\textit{Continued on next page}} \\
\endfoot
\hline
\endlastfoot

La V & \begin{tabular}{l} ${\bf 5p^{5}}$, $5p^{4} 4f^{1}$, $5p^{4} 6s^{1}$, $5p^{4} 5d^{1}$, $5p^{4} 6p^{1}$\end{tabular} & 89 & 1 124 & 1 124\\ 
La VI & \begin{tabular}{l} ${\bf 5p^{4}}$, $5p^{3} 4f^{1}$, $5p^{3} 6s^{1}$, $5p^{3} 5d^{1}$, $5p^{3} 6p^{1}$\end{tabular} & 121 & 1 937 & 1 937\\ 
La VII & \begin{tabular}{l} ${\bf 5p^{3}}$, $5p^{2} 4f^{1}$, $5p^{2} 6s^{1}$, $5p^{2} 5d^{1}$, $5p^{2} 6p^{1}$\end{tabular} & 92 & 1 208 & 1 208\\ 
La VIII & \begin{tabular}{l} ${\bf 5p^{2}}$, $5p^{1} 4f^{1}$, $5p^{1} 6s^{1}$, $5p^{1} 5d^{1}$, $5p^{1} 6p^{1}$\end{tabular} & 43 & 254 & 254\\ 
La IX & \begin{tabular}{l} ${\bf 5p^{1} 5s^{2}}$, $5p^{1} 5s^{1} 4f^{1}$, $5p^{1} 5s^{1} 6s^{1}$, $5p^{1} 5s^{1} 5d^{1}$,\\$5p^{1} 5s^{1} 6p^{1}$\end{tabular} & 74 & 808 & 808\\ 
La X & \begin{tabular}{l} ${\bf 5s^{2}}$, $5s^{1} 4f^{1}$, $5s^{1} 6s^{1}$, $5s^{1} 5d^{1}$, $5s^{1} 6p^{1}$\end{tabular} & 15 & 30 & 30\\
\end{longtable}
\end{center}

\end{document}